\documentclass[twocolumn]{aastex631}

\hypersetup{linkcolor=blue,citecolor=blue,filecolor=cyan,urlcolor=magenta}

\usepackage{tabularx}
\usepackage{url}
\usepackage{graphicx}
\usepackage[T1]{fontenc}
\usepackage{ae,aecompl}
\usepackage{booktabs}
\usepackage{natbib}
\usepackage{enumitem}
\usepackage{mathtools}
\usepackage{amssymb}
\usepackage{amsmath}
\usepackage{ulem}
\usepackage{epstopdf}
\usepackage{color}
\usepackage{xcolor}
\usepackage{needspace}
\usepackage{soul}

\usepackage{blindtext}
\usepackage{longtable}
\usepackage{tabu}

\usepackage{appendix}
\usepackage{rotating}

\usepackage{lipsum} 

\usepackage{blindtext}
\usepackage{longtable}
\usepackage{tabu}

\usepackage{appendix}

\usepackage{amsmath} 

\def\aap{A\&A} \def\apjl{ApJ}  
 \def\apjs{ApJS}

\shorttitle{exoplanet candidate prioritization with ML methods}

\begin{document}


\title{Towards Instrument-Agnostic Exoplanet Candidate Prioritization}

\author[0009-0002-4527-3865]{Vivaswan Kopparapu} 
\affiliation{Atholton High School, 6520 Freetown Rd, Columbia, MD 21044}
\affiliation{Astrophysics Science Division, NASA Goddard Space Flight Center, Greenbelt, MD 20771, USA.}

\author[0000-0003-2714-0487]{Sibasish Laha} 
\affiliation{Astrophysics Science Division, NASA Goddard Space Flight Center, Greenbelt, MD 20771, USA.}
\affiliation{Center for Space Science and Technology, University of Maryland Baltimore County, 1000 Hilltop Circle, Baltimore, MD 21250, USA.}
\affiliation{Center for Research and Exploration in Space Science and Technology, NASA/GSFC, Greenbelt, Maryland 20771, USA}

\author[0000-0000-0000-0000]{Brian P. Powell} 
\affiliation{Astrophysics Science Division, NASA Goddard Space Flight Center, Greenbelt, MD 20771, USA.}

\author[0000-0001-9786-1031]{Veselin Kostov} 
\affiliation{Astrophysics Science Division, NASA Goddard Space Flight Center, Greenbelt, MD 20771, USA.}
\affiliation{SETI Institute, 189 Bernardo Ave, Suite 200, Mountain View, CA 94043, USA}

\author[0009-0008-0344-0278]{Rajarshi Basak} 

\affiliation{Department of Computer Sciences, University of Maryland Baltimore County, 1000 Hilltop Circle Baltimore, MD 21250, USA}
\affiliation{Astrophysics Science Division, NASA Goddard Space Flight Center, Greenbelt, MD 20771, USA.}

\author[0009-0009-6408-5473]{Samuel Verbrugge} 
\affiliation{Astrophysics Science Division, NASA Goddard Space Flight Center, Greenbelt, MD 20771, USA.}
\affiliation{Department of Astronomy, University of Maryland College Park, 4296 Stadium Drive College Park, MD 20742, USA}


\begin{abstract}

We have developed a novel machine learning (ML) approach for predicting the likelihood of exoplanet candidate confirmation equally capable of performance on both TESS and Kepler data. From the NASA exoplanet archival post-processed Kepler and TESS databases, we chose six parameters that we assessed to be predictive to the planet transit signature: planet orbital period ({\it P}), planet radius ($R_{p}$), stellar temperature ($T_\text{eff}$), stellar radius ($R_{\star}$), planet transit depth ($\delta$), and planet transit duration ($t_{d}$).  We used these parameters to evaluate eleven different ML models on all possible train/test combinations of TESS and Kepler data, using the confirmed planet and false positive labels as our training targets. We found that, due to substantially different distributions of our chosen parameters in Kepler and TESS databases,  models trained with data from one instrument have difficulty predicting the other.  However, models trained jointly with both TESS and Kepler data can perform well on both.  We combined our best models into a statistically robust ensemble to evaluate the planet candidates in both Kepler and TESS, and we provide a list of the top candidates predicted by our model for each. Confirmed planets and false positives that have been resolved since the completion of our analysis demonstrate the effectiveness of our model and suggest that our top candidates are likely to be confirmed if they are further analyzed by the community.  With the upcoming launch of the Nancy Grace Roman Space Telescope (Roman) and the expected order-of-magnitude increase in planet candidates, we suggest that our method can be extended to Roman data for robust and effective prioritization for analysis.

\end{abstract}


\section{Introduction} \label{sec:intro}

The transit method, periodic dimming of a star as a planet crosses its disk, has delivered the majority of the $\sim$6,000 known exoplanets to date. The depth, duration, and morphology of a transit light curve provide the planet-to-star radius ratio, orbital period, impact parameter,  bulk density and composition constraints. NASA's \textit{Kepler} \citep{koch2010,borucki2010, Jenkins2010a} and \textit{Transiting Exoplanet Survey Satellite} \citep[TESS,][]{Ricker2015} have exploited this technique on a large scale, observing millions of stars, but with distinct survey strategies and data characteristics that yield complementary planet samples and demographics.

\textit{Kepler} was designed for a statistical census, continuously monitoring $\sim$150{,}000 Sun-like stars in a single field with high photometric stability and a long-cadence sampling of 29.4\,min (with 1\,min short cadence for selected targets). The mission enabled detections of shallow, long-period signals that approach Earth analog sensitivity. In fact, one of the primary scientific goals of the \textit {Kepler} prime mission was to estimate the occurrence of terrestrial planets in the habitable zones (HZs) of Sun-like stars. 
Following the loss of two reaction wheels, \textit{Kepler} was repurposed as the K2 mission,  pointing along the ecliptic and observed a sequence of $\sim$70--80\,day campaigns across $\gtrsim$20 fields \citep{Howell2014}. 
The mission yielded hundreds of validated and confirmed planets, including multi-planet systems and small planets around nearby M dwarfs such as K2-3 and K2-18, that are amenable to mass measurements and atmospheric follow-up \citep{Crossfield2015,Montet2015}. 

In contrast, TESS has surveyed nearly the entire sky in $\sim$27-day sectors for over seven years,
prioritizing bright, nearby stars. During the two-year prime mission, TESS recorded 2\,min cadence selected targets and 30\,min cadence full-frame images (FFIs); in the extended missions, TESS added 20\,s cadence and shortened FFI cadence to 10\,min and then 200\,s, expanding sensitivity to short-period and small-radius planets. 

Together these missions allowed for estimating the occurrence rate of planets among diverse stellar spectral types. Data from the \textit{Kepler} mission indicated that small planets are common around FGK stars, with occurrence rate of HZ planets $\eta_\oplus \sim 0.37^{+0.48}_{-0.21}$\citep{Bryson2021}\footnote{{\bf Note that other values are provided by \citet{Burke2015}, \citet{Hsu2019}, and \citet{Kunimoto2020}}}, and $0.0858^{+0.179}_{-0.0822}$ for M-dwarfs ($T_\text{eff}< 4000$K) \citep{bergsten2023}. TESS has a different bandpass than Kepler in order to facilitate observations of planets around M dwarfs. Early TESS demographic work has measured short-period planet rates for terrestrial planets around nearby mid-to-late M dwarfs with orbital periods $< 7$ days and found that terrestrials outnumber sub-Neptunes around by 14 to 1 \citep{ment2023a}.

\begin{figure*}
    \centering
    \includegraphics[width=12cm]{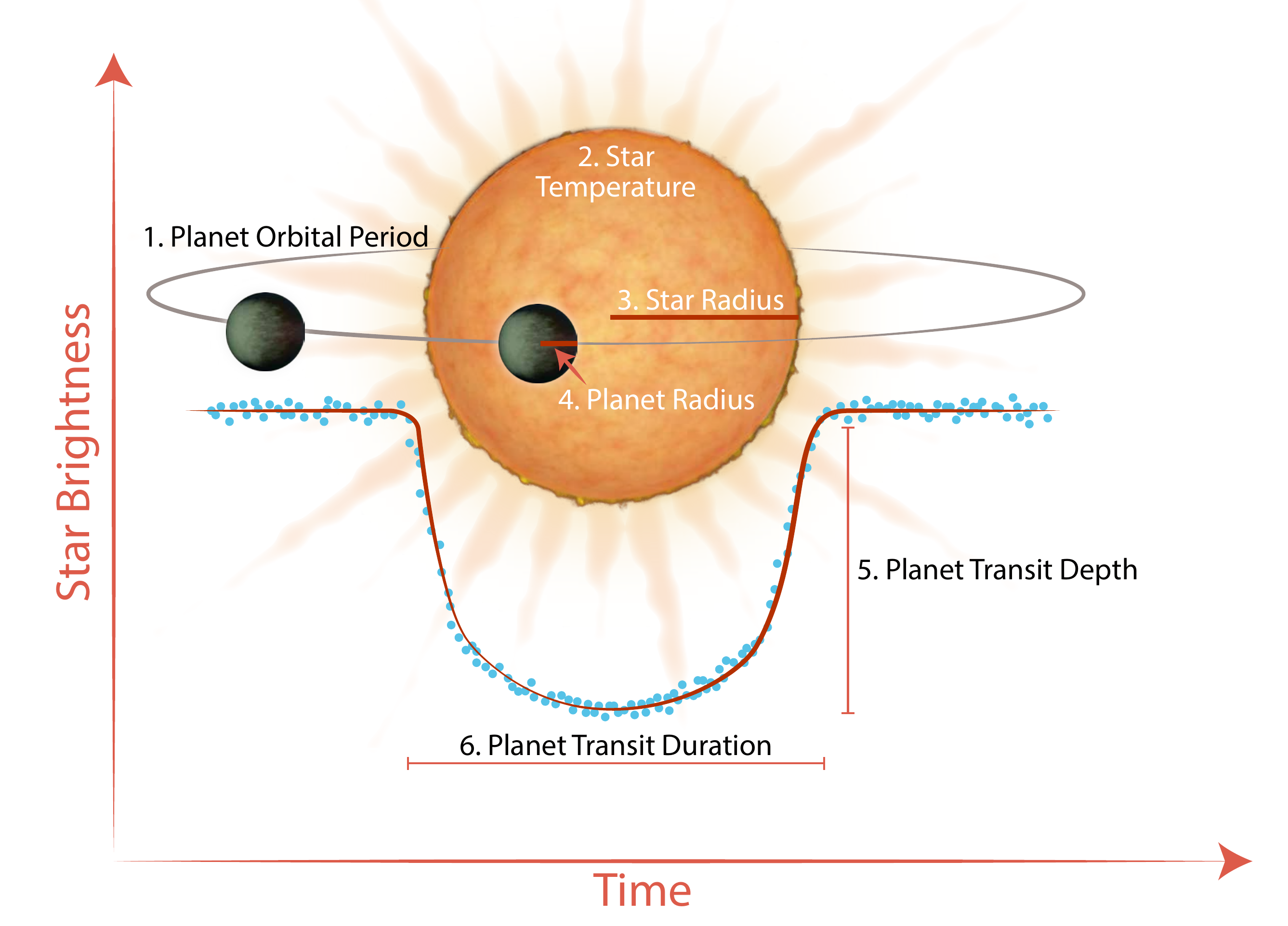}
    \caption{Schematic showing a planet transiting a star. The six parameters that we used from the TESS and Kepler databases to train our ML models are noted in the diagram.}
    \label{fig::planet_transition}
\end{figure*}

All of the planetary detections in TESS and Kepler have undergone extensive manual and automated vetting techniques. In particular, identifying the false positives (FP) among the long list of planetary candidates in either databases is an important step. Stellar variability and instrument noise sometimes mimic a planet-transition, leading to FPs.  Several automated vetting algorithms and automated-pipelines have also been developed for that purpose, for example, Robovetter \citep{2018ApJS..235...38T}, or, more recently LEOvetter \citep{Kunimoto_2025}, RAVEN \citep{2025arXiv250917645H} and AstroNet \citep{astronet}, among many others. Although the methodology and approach for vetting may differ from algorithm to algorithm, the objective generally remains to confirm or deny the presence of an exoplanet based on characteristics of the transit and host star, whether at the raw pixel level and/or various stages of post-processing.

{Existing literature regarding the use of machine learning (ML) for exoplanet transit classification is trending toward sophisticated deep learning architectures with a wealth of inputs. Astronet-Triage-v2 \citep{2023AJ....165...95T}, for example, uses a convolutional neural network (CNN) that processes each transit crossing event (TCE) through a multitude of phase-folded views.  More recently, ExoMiner++ \citep{2025AJ....170..287V} comprehensively combines the full suite of data validation products in a multi-branch CNN.  In pursuing statistical validation of planet candidates, \citet{2021MNRAS.504.5327A} provides an ensemble of classification models in a Bayesian validation framework using 38 Kepler pipeline diagnostics.}  

{ In this work, we detail a complementary method to the comprehensive classification and validation approaches described above that emphasizes simplicity and generalizability for a less complex problem that yields a practical outcome: Prioritization.  As the number of planet candidates grows, a means of intelligently selecting candidates for further in-depth analysis is of paramount importance.  To this end, we provide a novel contribution to the ongoing research and improvement of planet-vetting techniques in the ability to generalize models to be instrument-agnostic in predicting the likelihood of confirmation among planetary candidates in both the TESS and Kepler planet databases.  We use only six features from the post-processed TESS and Kepler planet transit data in the NASA Exoplanet Archive\footnote{\url{https://exoplanetarchive.ipac.caltech.edu/index.html}} to train various ML models to accurately make this prediction.}  By training our models on TESS and Kepler data jointly using standard post-processed parameters, we create a { simple}, robust, and generalizable model that can be applied to current and future missions.  We also provide predictions for the planetary candidates from both TESS and Kepler to assist in prioritization of candidates for follow-up and confirmation efforts.    

This paper is arranged as follows: In Section \ref{sec::data}, we discuss the TESS and Kepler databases, followed by Section \ref{sec::ML} where we discuss the ML models and data preparation. In Section \ref{sec::results} we provide the data distributions, the training methodology, various model results, and our predictions for planetary candidates.  Lastly, we provide a discussion and {summary} in Section \ref{sec::discussion} and Section \ref{sec::summary}, respectively.

\section{TESS and Kepler Data }
\label{sec::data}

We obtained the TESS and the Kepler databases  from the NASA Exoplanet Science Institute (NExScI)\footnote{\url{https://nexsci.caltech.edu/}}. 
Details regarding how these databases are curated by NExScI can be found on their separate websites for TESS\footnote{\url{https://exoplanetarchive.ipac.caltech.edu/docs/TESSMission.html}} and Kepler\footnote{\url{https://exoplanetarchive.ipac.caltech.edu/docs/Kepler_KOI_docs.html}}. 

For the TESS dataset, we selected the TESS TOI catalog (current as of August 30, 2025) that includes false positives (FP), planet candidates (PC), ambiguous planetary candidate (APC), false alarm (FA), confirmed planet (CP), and known planet (KP). For the Kepler database, {we used the cumulative KOI catalog (current as of September 28, 2025), which included false positives, confirmed-planets, and candidates}.   We grouped the TESS FP and FA into a combined FP category and CP and KP into a combined CP category.  We therefore have three clear classes in both lists: CP, FP and candidates. The CP and FPs form the basis of our labeled training and testing datasets for the machine learning models which are then applied to unlabeled candidates. {In the TESS database we have a fairly balanced number of CPs (1198) and FPs (1137). { Whereas, in the Kepler database,} there are 2744 CPs, 4582 FPs, and 1874 PCs. }

\begin{figure*}
    \centering
    \includegraphics[width=\textwidth]{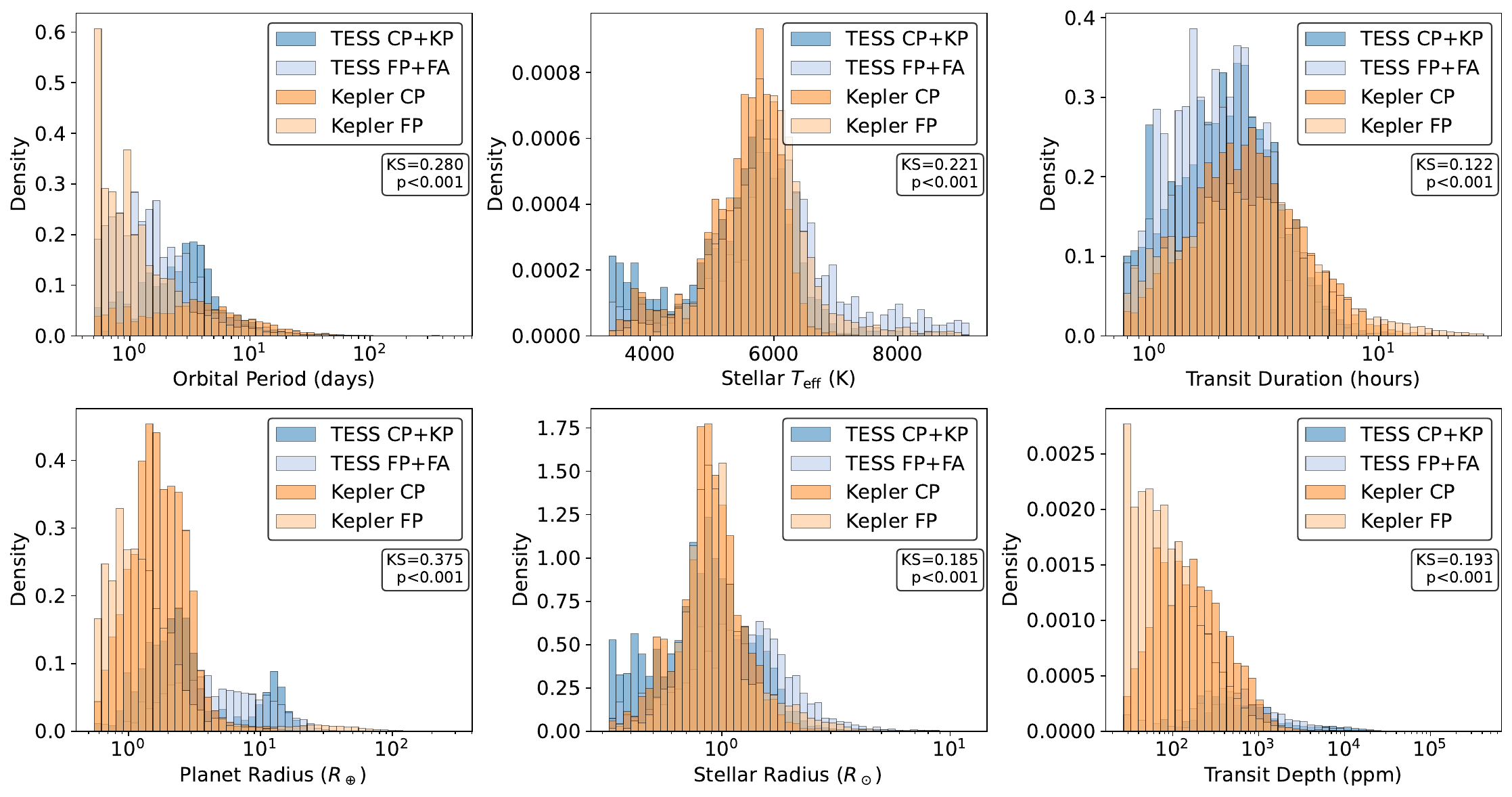}
    \caption{Distribution of input features separated by source (TESS/Kepler) and label (confirmed planets/false positives). TESS CP+KP (n=1198, dark blue), TESS FA+FP (n=1137, light blue), Kepler CP (n=2744, dark orange), Kepler FP (n=4582, light orange). All features except T$_\text{eff}$ are plotted on a logarithmic x-axis. KS test results between the full distributions (i.e. all TESS and all Kepler) are noted in a box below the legend.  Differences between { CP and FP} distributions within each survey indicate the discriminative power of each feature, while differences between surveys highlight domain shift challenges.}
    \label{fig::dist}
\end{figure*}

From both the TESS and Kepler lists, we selected six post-processed data columns for our analysis: planet orbital period ({\it P}), planet radius ($R_{p}$), stellar temperature ($T_\text{eff}$), stellar radius ($R_{\star}$), planet transit depth ($\delta$), and planet transit duration ($t_{d}$).  These parameters are shown in a helpful schematic in Figure \ref{fig::planet_transition}, with distributions of each parameter given in Figure \ref{fig::dist} (we will discuss these distributions further in Section \ref{sec::results}).  We acknowledge inter-dependencies among these parameters. For example, the transit depth depends upon the radius of the planet and the radius of the star, and can be approximated through the equation
\begin{equation}
   \delta  \approx \biggl(\frac{R_{p}}{R_{\star}}\biggr)^{2}
\end{equation}
Similarly, the transit duration can be approximated using the orbital parameters of the planet and the radius of the star
\begin{equation}
t_{d}\approx \frac{P}{\pi a}~R_{\star},
\end{equation}
where $a$ is the semi-major axis of the orbit.  

We also notably excluded magnitudes from our analysis because magnitude is a combination of stellar luminosity and the squared distance of the star from Earth, the latter of which is not a valid quantity in determining the presence/absence of an exoplanet and the former is redundant as a proxy for $T_\text{eff}$, which is already a selected feature.  {Furthermore, bandpass differences between observatories make magnitudes difficult to compare in an instrument-agnostic framework.  The signal-to-noise ratio carries information about magnitude and would be an excellent input feature for this effort if it were uniformly catalogued, which it unfortunately is not.}

The post-processed parameters in the TESS and Kepler databases have been extensively human-vetted and may have intrinsic biases, for example, in favor of M-dwarf stars (being more abundant) or brighter stars (better signal-to-noise ratio).  { Planet candidates that were subsequently labeled CP or FP} have been so because they possessed characteristics which which were amenable to classification. We acknowledge this bias and clearly assert that the intent of our model is not to be unbiased, rather it is for the practical value of learning relationships between input quantities that allow for binary classification as CP or FP.  In this manner, bias is a helpful attribute rather than a statistical hindrance.


\section{Machine Learning Models and Data Preparation}
\label{sec::ML}

Our intent with this work is to determine one or more ML models that can effectively train on the labeled (CP and FP) TESS and Kepler datasets with the six chosen parameters of the post-processed data and accurately predict the presence/absence of an exoplanet from the unlabeled data (i.e.,~to \textit{generalize}). Extending this further, we seek to develop a model capable of being instrument-agnostic, that is, to perform equally well on both TESS and Kepler data.  To that end, we have employed eleven common supervised ML models in our experiment in order to determine which model (and under what circumstances) yields the best predictive power. The models are: AdaBoost (AB), Decision Tree (DT), Gaussian Naive Bayes (GNB), Gradient Boosting (GB), K-Nearest Neighbors (KNN), Multi-Layer Perceptron (MLP), Quadratic Discriminant Analysis (QDA), Random Forest (RF), Support Vector Machines (SVM), LightGBM (LGBM), and XGBoost (XGB). All models except the last two noted are readily available in Python {\sc scikit-learn} library \citep{pedregosa2011scikit}. The XGBoost \citep{chen2016xgboost} and LightGBM \citep{ke2017lightgbm} models are available through the {\sc xgboost} and {\sc lightgbm} python libraries, respectively. These models are ideal for our experiments because they work well with the size of our data and provide a wide variety of methodologies, increasing the likelihood that we will find a method with strong predictive power yet capable of broad generalization.  

Of our six selected input parameters, some have ranges that are orders of magnitude apart, e.g. $P$ ranges from 0.16 days to 1,825 days.  In such cases, many types of ML models will fail to train without adequate scaling performed in pre-processing. The best example of this is the MLP, which cannot viably train without similarly-scaled values due to the inherent limitations of neural networks. In contrast, tree-based methods such as RF do not need preliminary scaling, as the algorithm is fundamentally a learned series of nested if-statements, {comparing boolean derivations between quantities rather than numerical values}. For a thorough comparison of the models as applied to our data, we trained each {scale-sensitive} model {(KNN, MLP, and SVC)} with both scaled and unscaled data, using the \texttt{MinMax} scaler as provided by {\sc scikit-learn}, which rescales the values of each feature to a [0,1] range.  {For models which were not scale-sensitive (all others), we used only unscaled data.}

\section{Results}
\label{sec::results}

Having both Kepler and TESS datasets available to us, and with a goal of developing the best and most generalizable model, we trained and tested in all possible combinations of the two datasets, shown for clarity with abbreviations for train/test configurations used hereafter in Table \ref{tab::train_test}. 

Despite the fact that the TESS and Kepler feature sets are all post-processed, it is easy to see in Figure \ref{fig::dist} that they have very different distributions between the instruments.  We show these distributions for CPs and FPs and also note the Kolmogorov–Smirnov (KS) test results between the combined distributions for TESS and Kepler on each plot.  The null hypothesis of the KS test is that the distributions are identical, with a small p-value (generally $p<0.05$) indicating a rejection of the null hypothesis.  KS test values give the maximum difference between the distribution CDFs, or $D$-statistic, while the $p$ value gives the probability of observing the $D$-statistic if the null hypothesis is true. In summary, the test indicates that the distributions for {\it all} of our input features are very different between TESS and Kepler.  As related to our attempts to train on one dataset and test on the other, the KS tests suggested that we should anticipate difficulty.

{ It is clear from Figure \ref{fig::dist} that TESS and Kepler observed different populations {of} planet and stellar characteristics. Though out of scope for us to examine these differences closely and speculate on their relative causes, {we note that extensive analysis of the Kepler transiting exoplanet population has been conducted by \citet{2013ApJ...766...81F}, \citet{2018ApJS..235...38T}, and \citet{2020AJ....160..108B}, and of the TESS transiting exoplanet population by \citet{2021ApJS..254...39G}.}  As it relates specifically to our analysis, it is important to consider that { training exclusively on} the different distributions will lead to ML models that are highly disparate between the two missions.}

In the remainder of this section, we will evaluate the performance of each of our models trained on TESS, Kepler, and the combination of both.  All our tables will show the six performance metrics, where we highlight in bold the results of models where all six of these metrics are $\ge 0.75$, {which { we set as} an arbitrary cutoff to demonstrate high-performing models}.

\begin{table}[h]
\centering
\caption{Train/Test Configuration Abbreviations}
\begin{tabular}{|l|c|c|c|}
\hline
Training Data & \multicolumn{3}{c|}{Testing Data} \\ \cline{2-4}
 & TESS & Kepler & TESS+Kepler \\ \hline
TESS & T/T & T/K & T/TK \\ \hline
Kepler & K/T & K/K & K/TK \\ \hline
TESS+Kepler & TK/T & TK/K & TK/TK \\ \hline
\end{tabular}
\label{tab::train_test}
\end{table}

\begin{figure*}[htp]
    \centering
    \includegraphics[width=15cm,height=18cm]{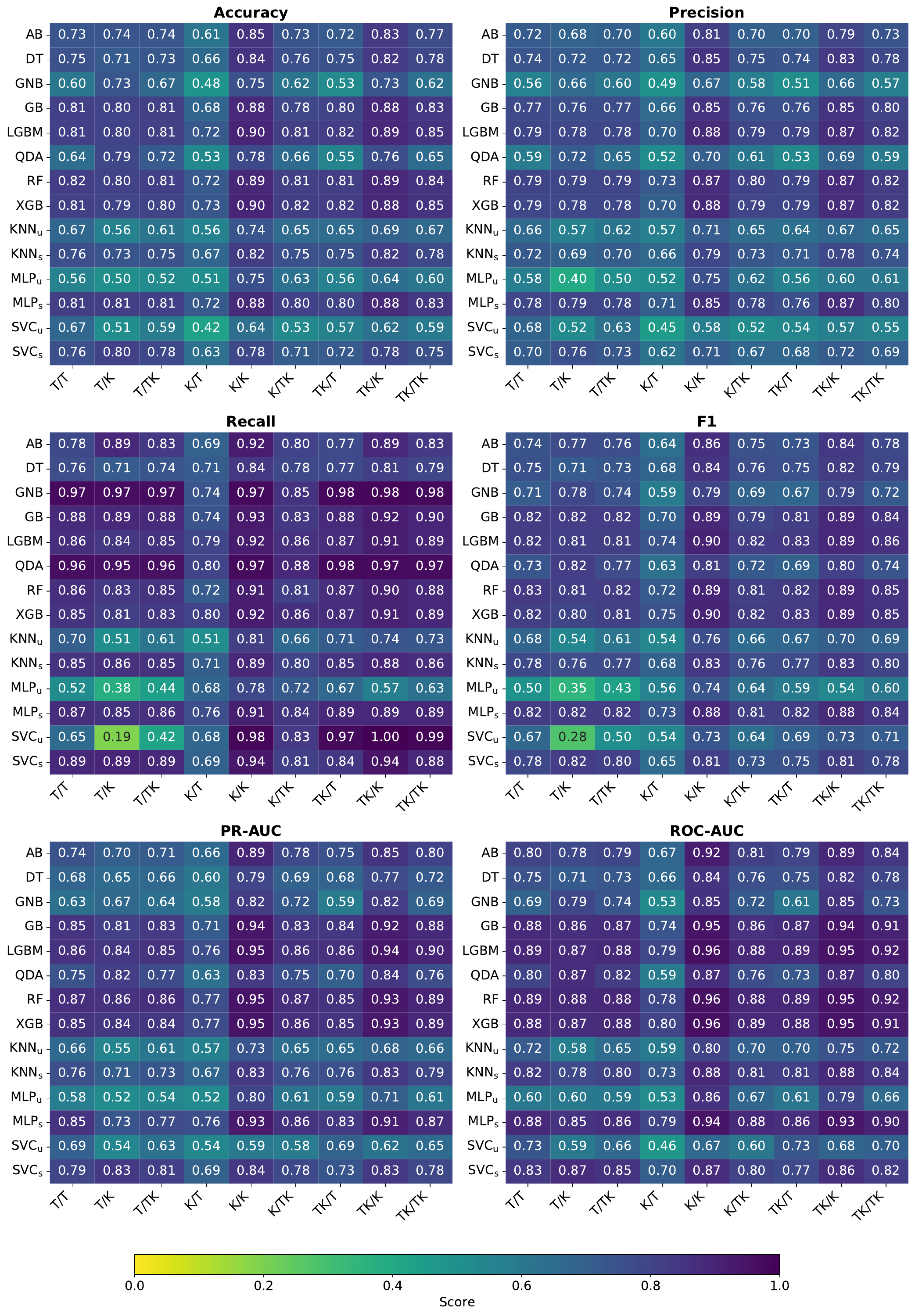}
    \caption{Heatmaps showing mean classifier performance across all train/test configurations using 20-iteration repeated random sub-sampling. Color scale ranges from 0.0 (red) to 1.0 (green). Panels show (top-left to bottom-right): Accuracy, Precision, Recall, F1, PR-AUC, and ROC-AUC.  For scale-sensitive models, subscripts `u' and `s' indicate unscaled and scaled, respectively.  Train/test configuration abbreviations are given in Table \ref{tab::train_test}.}
    \label{fig::heatmaps}
\end{figure*}

\subsection{Training Methodology}
\label{sec::methodology}

For the purposes of confirming model fidelity, a common choice { for cross-validation} is a five-fold analysis, whereby a dataset is split into five groups each containing 20\% of the data.  The model is then trained and tested five times, each time using a different group of 20\% as the test set with the remaining 80\% as the training set.  An additional consideration is the necessity of a balanced training and testing set, with 50\% each of positives (CP) and negatives (FP) such that the model does not develop a bias toward prediction one way or the other.  In our particular application, since we combined TESS and Kepler data, this requirement presents a substantial obstacle (recall from Section \ref{sec::data} that TESS has 1,198 positives and 1,137 negatives, whereas the Kepler data contains 2,744 positives and 4,582 negatives).  Yet another consideration is that, in pursuit of generalizability, we needed to ensure that {models trained on both TESS and Kepler data do not favor the distinct characteristics of either data set}.  As such, we had two separate balancing requirements:  label (positive/negative) and source (TESS/Kepler).  A properly-conducted five-fold analysis, then, would require excluding the majority of our data.  This was, of course, unacceptable.

We chose instead to randomly resample our data.  This is another common method of cross-validation (see, e.g.~{\sc scikit-learn}'s \texttt{StratifiedShuffleSplit} function), originally described by \citet{kohavi1995study}.  In each case of our model training (described in Sections \ref{sec::tess_train}, \ref{sec::kepler_train}, and \ref{sec::combined_train}) we randomly resampled a balanced (by label and source) subset for training and testing in a 80/20 ratio, performing 20 iterations.  Our reported results are an average over these iterations, ensuring that our model comparisons and eventual predictions (see Section \ref{sec::predictions}) are statistically robust.

We show the complete results of our assessments in Figure \ref{fig::heatmaps}, { with six heatmap-style plots showing the accuracy (top left), precision (top right), recall (middle left), F1 score (middle right), PR-AUC (bottom left) and ROC-AUC (bottom right).}  The text in each box shows the {average} numerical score, while the color scale shows zero to one on a { perceptually-uniform gradient of yellow to purple, where darker is better}.  We note that precision and recall are complementary, such that a high score in one without a similarly high score in the other does not indicate a successful model.  For example, the QDA and GNB recall scores are near-universally excellent, while their precision scores are near-universally poor.  This performance indicates that the model tended to predict high values in nearly all cases, which allows for excellent recall but poor precision.  An effective and balanced model will have both high precision and high recall, reflected in the F1 score, which is the harmonic mean of the two.  { The most informative individual metrics to describe model performance in our case are the PR-AUC and ROC-AUC, which reflect performance across the full range of numerical predictions.}

\subsection{Training on TESS Data}
\label{sec::tess_train}

\begin{table*}[htb]
\centering
\caption{Model performance: Train on TESS, Test on TESS (T/T), {\bf values reported as $\mu\pm\sigma$}}
\label{tab::t_t}
\resizebox{\textwidth}{!}{%
\begin{tabular}{lccccccc}
\toprule
Model & Sc. & Acc & Prec & Rec & F1 & PR-AUC & ROC-AUC \\
\midrule
\multicolumn{8}{l}{\textit{Scale-Invariant Models}} \\
\midrule
AB & N/A & 0.73$\pm$0.02 & 0.72$\pm$0.02 & 0.78$\pm$0.04 & 0.74$\pm$0.02 & 0.74$\pm$0.02 & 0.80$\pm$0.02 \\
DT & N/A & 0.75$\pm$0.02 & 0.74$\pm$0.02 & 0.76$\pm$0.03 & 0.75$\pm$0.02 & 0.68$\pm$0.02 & 0.75$\pm$0.02 \\
GNB & N/A & 0.60$\pm$0.01 & 0.56$\pm$0.01 & 0.97$\pm$0.01 & 0.71$\pm$0.01 & 0.63$\pm$0.03 & 0.69$\pm$0.02 \\
{\bf GB} & {\bf N/A} & {\bf 0.81$\pm$0.01} & {\bf 0.77$\pm$0.01} & {\bf 0.88$\pm$0.01} & {\bf 0.82$\pm$0.01} & {\bf 0.85$\pm$0.02} & {\bf 0.88$\pm$0.01} \\
{\bf LGBM} & {\bf N/A} & {\bf 0.81$\pm$0.01} & {\bf 0.79$\pm$0.02} & {\bf 0.86$\pm$0.02} & {\bf 0.82$\pm$0.01} & {\bf 0.86$\pm$0.02} & {\bf 0.89$\pm$0.01} \\
QDA & N/A & 0.64$\pm$0.01 & 0.59$\pm$0.01 & 0.96$\pm$0.01 & 0.73$\pm$0.01 & 0.75$\pm$0.03 & 0.80$\pm$0.02 \\
{\bf RF} & {\bf N/A} & {\bf 0.82$\pm$0.01} & {\bf 0.79$\pm$0.01} & {\bf 0.86$\pm$0.02} & {\bf 0.83$\pm$0.01} & {\bf 0.87$\pm$0.02} & {\bf 0.89$\pm$0.01} \\
{\bf XGB} & {\bf N/A} & {\bf 0.81$\pm$0.01} & {\bf 0.79$\pm$0.02} & {\bf 0.85$\pm$0.02} & {\bf 0.82$\pm$0.01} & {\bf 0.85$\pm$0.02} & {\bf 0.88$\pm$0.01} \\
\midrule
\multicolumn{8}{l}{\textit{Scale-Sensitive Models}} \\
\midrule
KNN & N & 0.67$\pm$0.02 & 0.66$\pm$0.02 & 0.70$\pm$0.02 & 0.68$\pm$0.02 & 0.66$\pm$0.02 & 0.72$\pm$0.02 \\
KNN & Y & 0.76$\pm$0.02 & 0.72$\pm$0.02 & 0.85$\pm$0.03 & 0.78$\pm$0.02 & 0.76$\pm$0.02 & 0.82$\pm$0.01 \\
MLP & N & 0.56$\pm$0.05 & 0.58$\pm$0.07 & 0.52$\pm$0.27 & 0.50$\pm$0.19 & 0.58$\pm$0.07 & 0.60$\pm$0.10 \\
{\bf MLP} & {\bf Y} & {\bf 0.81$\pm$0.01} & {\bf 0.78$\pm$0.02} & {\bf 0.87$\pm$0.02} & {\bf 0.82$\pm$0.01} & {\bf 0.85$\pm$0.02} & {\bf 0.88$\pm$0.01} \\
SVC & N & 0.67$\pm$0.02 & 0.68$\pm$0.02 & 0.65$\pm$0.03 & 0.67$\pm$0.02 & 0.69$\pm$0.03 & 0.73$\pm$0.02 \\
SVC & Y & 0.76$\pm$0.02 & 0.70$\pm$0.02 & 0.89$\pm$0.02 & 0.78$\pm$0.01 & 0.79$\pm$0.02 & 0.83$\pm$0.02 \\
\bottomrule
\end{tabular}%
}
\small
Models: AB=AdaBoost, DT=Decision Tree, GB=Gradient Boosting,
GNB=Gaussian Naive Bayes, KNN=K-Nearest Neighbors, LGBM=LightGBM, MLP=Multi-Layer Perceptron,
QDA=Quadratic Discriminant Analysis, RF=Random Forest, SVC=Support Vector Classifier (RBF),
XGB=XGBoost. Sc.=Scaling, N/A=not applicable, N=No, Y=Yes.
Bold rows: all metrics $>$ 0.75. Values: mean $\pm$ std over 20 iterations with random resampling.
\end{table*}

\begin{table*}[htb]
\centering
\caption{Model performance: Train on TESS, Test on Kepler (T/K), {\bf values reported as $\mu\pm\sigma$}}
\label{tab::t_k}
\resizebox{\textwidth}{!}{%
\begin{tabular}{lccccccc}
\toprule
Model & Sc. & Acc & Prec & Rec & F1 & PR-AUC & ROC-AUC \\
\midrule
\multicolumn{8}{l}{\textit{Scale-Invariant Models}} \\
\midrule
AB & N/A & 0.74$\pm$0.02 & 0.68$\pm$0.02 & 0.89$\pm$0.02 & 0.77$\pm$0.02 & 0.70$\pm$0.03 & 0.78$\pm$0.03 \\
DT & N/A & 0.71$\pm$0.03 & 0.72$\pm$0.04 & 0.71$\pm$0.04 & 0.71$\pm$0.03 & 0.65$\pm$0.03 & 0.71$\pm$0.03 \\
GNB & N/A & 0.73$\pm$0.01 & 0.66$\pm$0.01 & 0.97$\pm$0.02 & 0.78$\pm$0.01 & 0.67$\pm$0.02 & 0.79$\pm$0.02 \\
{\bf GB} & {\bf N/A} & {\bf 0.80$\pm$0.01} & {\bf 0.76$\pm$0.02} & {\bf 0.89$\pm$0.03} & {\bf 0.82$\pm$0.01} & {\bf 0.81$\pm$0.03} & {\bf 0.86$\pm$0.02} \\
{\bf LGBM} & {\bf N/A} & {\bf 0.80$\pm$0.02} & {\bf 0.78$\pm$0.02} & {\bf 0.84$\pm$0.03} & {\bf 0.81$\pm$0.02} & {\bf 0.84$\pm$0.02} & {\bf 0.87$\pm$0.02} \\
QDA & N/A & 0.79$\pm$0.01 & 0.72$\pm$0.01 & 0.95$\pm$0.02 & 0.82$\pm$0.01 & 0.82$\pm$0.04 & 0.87$\pm$0.02 \\
{\bf RF} & {\bf N/A} & {\bf 0.80$\pm$0.01} & {\bf 0.79$\pm$0.02} & {\bf 0.83$\pm$0.04} & {\bf 0.81$\pm$0.02} & {\bf 0.86$\pm$0.01} & {\bf 0.88$\pm$0.01} \\
{\bf XGB} & {\bf N/A} & {\bf 0.79$\pm$0.01} & {\bf 0.78$\pm$0.02} & {\bf 0.81$\pm$0.04} & {\bf 0.80$\pm$0.02} & {\bf 0.84$\pm$0.02} & {\bf 0.87$\pm$0.01} \\
\midrule
\multicolumn{8}{l}{\textit{Scale-Sensitive Models}} \\
\midrule
KNN & N & 0.56$\pm$0.02 & 0.57$\pm$0.02 & 0.51$\pm$0.04 & 0.54$\pm$0.03 & 0.55$\pm$0.01 & 0.58$\pm$0.02 \\
KNN & Y & 0.73$\pm$0.02 & 0.69$\pm$0.02 & 0.86$\pm$0.02 & 0.76$\pm$0.01 & 0.71$\pm$0.02 & 0.78$\pm$0.02 \\
MLP & N & 0.50$\pm$0.14 & 0.40$\pm$0.21 & 0.38$\pm$0.37 & 0.35$\pm$0.29 & 0.52$\pm$0.08 & 0.60$\pm$0.12 \\
MLP & Y & 0.81$\pm$0.02 & 0.79$\pm$0.02 & 0.85$\pm$0.02 & 0.82$\pm$0.01 & 0.73$\pm$0.03 & 0.85$\pm$0.02 \\
SVC & N & 0.51$\pm$0.01 & 0.52$\pm$0.03 & 0.19$\pm$0.02 & 0.28$\pm$0.02 & 0.54$\pm$0.02 & 0.59$\pm$0.03 \\
{\bf SVC} & {\bf Y} & {\bf 0.80$\pm$0.01} & {\bf 0.76$\pm$0.02} & {\bf 0.89$\pm$0.01} & {\bf 0.82$\pm$0.01} & {\bf 0.83$\pm$0.02} & {\bf 0.87$\pm$0.01} \\
\bottomrule
\end{tabular}%
}

\small
 Models: AB=AdaBoost, DT=Decision Tree, GB=Gradient Boosting,
GNB=Gaussian Naive Bayes, KNN=K-Nearest Neighbors, LGBM=LightGBM, MLP=Multi-Layer Perceptron,
QDA=Quadratic Discriminant Analysis, RF=Random Forest, SVC=Support Vector Classifier (RBF),
XGB=XGBoost. Sc.=Scaling, N/A=not applicable, N=No, Y=Yes.
Bold rows: all metrics $>$ 0.75. Values: mean $\pm$ std over 20 iterations with random resampling.

\end{table*}

\begin{table*}[htb]
\centering
\caption{Model performance: Train on TESS, Test on TESS+Kepler (T/TK), {\bf values reported as $\mu\pm\sigma$}}
\label{tab::t_tk}
\resizebox{\textwidth}{!}{%
\begin{tabular}{lccccccc}
\toprule
Model & Sc. & Acc & Prec & Rec & F1 & PR-AUC & ROC-AUC \\
\midrule
\multicolumn{8}{l}{\textit{Scale-Invariant Models}} \\
\midrule
AB & N/A & 0.74$\pm$0.01 & 0.70$\pm$0.01 & 0.83$\pm$0.02 & 0.76$\pm$0.01 & 0.71$\pm$0.02 & 0.79$\pm$0.01 \\
DT & N/A & 0.73$\pm$0.02 & 0.72$\pm$0.02 & 0.74$\pm$0.03 & 0.73$\pm$0.02 & 0.66$\pm$0.02 & 0.73$\pm$0.02 \\
GNB & N/A & 0.67$\pm$0.01 & 0.60$\pm$0.01 & 0.97$\pm$0.02 & 0.74$\pm$0.01 & 0.64$\pm$0.02 & 0.74$\pm$0.02 \\
{\bf GB} & {\bf N/A} & {\bf 0.81$\pm$0.01} & {\bf 0.77$\pm$0.01} & {\bf 0.88$\pm$0.02} & {\bf 0.82$\pm$0.01} & {\bf 0.83$\pm$0.02} & {\bf 0.87$\pm$0.01} \\
{\bf LGBM} & {\bf N/A} & {\bf 0.81$\pm$0.01} & {\bf 0.78$\pm$0.01} & {\bf 0.85$\pm$0.02} & {\bf 0.81$\pm$0.01} & {\bf 0.85$\pm$0.01} & {\bf 0.88$\pm$0.01} \\
QDA & N/A & 0.72$\pm$0.01 & 0.65$\pm$0.01 & 0.96$\pm$0.02 & 0.77$\pm$0.01 & 0.77$\pm$0.03 & 0.82$\pm$0.01 \\
{\bf RF} & {\bf N/A} & {\bf 0.81$\pm$0.01} & {\bf 0.79$\pm$0.01} & {\bf 0.85$\pm$0.02} & {\bf 0.82$\pm$0.01} & {\bf 0.86$\pm$0.01} & {\bf 0.88$\pm$0.01} \\
{\bf XGB} & {\bf N/A} & {\bf 0.80$\pm$0.01} & {\bf 0.78$\pm$0.02} & {\bf 0.83$\pm$0.02} & {\bf 0.81$\pm$0.01} & {\bf 0.84$\pm$0.01} & {\bf 0.88$\pm$0.01} \\
\midrule
\multicolumn{8}{l}{\textit{Scale-Sensitive Models}} \\
\midrule
KNN & N & 0.61$\pm$0.01 & 0.62$\pm$0.02 & 0.61$\pm$0.02 & 0.61$\pm$0.02 & 0.61$\pm$0.01 & 0.65$\pm$0.02 \\
KNN & Y & 0.75$\pm$0.01 & 0.70$\pm$0.01 & 0.85$\pm$0.02 & 0.77$\pm$0.01 & 0.73$\pm$0.01 & 0.80$\pm$0.01 \\
MLP & N & 0.52$\pm$0.05 & 0.50$\pm$0.07 & 0.44$\pm$0.30 & 0.43$\pm$0.20 & 0.54$\pm$0.05 & 0.59$\pm$0.07 \\
{\bf MLP} & {\bf Y} & {\bf 0.81$\pm$0.01} & {\bf 0.78$\pm$0.02} & {\bf 0.86$\pm$0.02} & {\bf 0.82$\pm$0.01} & {\bf 0.77$\pm$0.03} & {\bf 0.86$\pm$0.02} \\
SVC & N & 0.59$\pm$0.01 & 0.63$\pm$0.02 & 0.42$\pm$0.02 & 0.50$\pm$0.02 & 0.63$\pm$0.01 & 0.66$\pm$0.02 \\
SVC & Y & 0.78$\pm$0.01 & 0.73$\pm$0.01 & 0.89$\pm$0.01 & 0.80$\pm$0.01 & 0.81$\pm$0.02 & 0.85$\pm$0.01 \\
\bottomrule
\end{tabular}%
}

\small
 Models: AB=AdaBoost, DT=Decision Tree, GB=Gradient Boosting,
GNB=Gaussian Naive Bayes, KNN=K-Nearest Neighbors, LGBM=LightGBM, MLP=Multi-Layer Perceptron,
QDA=Quadratic Discriminant Analysis, RF=Random Forest, SVC=Support Vector Classifier (RBF),
XGB=XGBoost. Sc.=Scaling, N/A=not applicable, N=No, Y=Yes.
Bold rows: all metrics $>$ 0.75. Values: mean $\pm$ std over 20 iterations with random resampling.

\end{table*}

In this section we discuss models trained on only the TESS data (T/T, T/K, T/TK, with results provided in Tables \ref{tab::t_t}, \ref{tab::t_k}, and \ref{tab::t_tk}, respectively).  {We highlighted in bold those models which met our success criteria of all six metrics $>0.75$.}  Starting a trend that the rest of our results would follow, we found that all the tree-based methods (GB, LGBM, RF, and XGB) performed very well.  The scaled MLP was the only scale-sensitive method to meet our success criteria { for T/T, T/K, and T/TK}.  Of course, as we discussed in Section \ref{sec::ML}, data scaling is a prerequisite for use of neural networks in nearly all circumstances.  Figure \ref{fig::heatmaps} demonstrates rather clearly how, in each of the scale-sensitive models, the models trained on unscaled data performed dismally.

We note that the results of training on TESS and testing on Kepler (scenario T/K, Table \ref{tab::t_k}), were better than we expected, and only moderately reduced from testing on TESS (scenario T/T, Table \ref{tab::t_t}).  In Section \ref{sec::kepler_train}, we will discuss this result as well as the inverse scenario (K/T, Table \ref{tab::k_t}).  { Testing on the combined TESS and Kepler data (scenario T/TK, Table \ref{tab::t_tk}) showed similar results to the T/K scenario.}

\subsection{Training on Kepler Data}
\label{sec::kepler_train}

\begin{table*}[htb]
\centering
\caption{Model performance: Train on Kepler, Test on TESS (K/T), {\bf values reported as $\mu\pm\sigma$}}
\label{tab::k_t}
\resizebox{\textwidth}{!}{%
\begin{tabular}{lccccccc}
\toprule
Model & Sc. & Acc & Prec & Rec & F1 & PR-AUC & ROC-AUC \\
\midrule
\multicolumn{8}{l}{\textit{Scale-Invariant Models}} \\
\midrule
AB & N/A & 0.61$\pm$0.02 & 0.60$\pm$0.02 & 0.69$\pm$0.04 & 0.64$\pm$0.02 & 0.66$\pm$0.03 & 0.67$\pm$0.03 \\
DT & N/A & 0.66$\pm$0.03 & 0.65$\pm$0.03 & 0.71$\pm$0.04 & 0.68$\pm$0.03 & 0.60$\pm$0.02 & 0.66$\pm$0.03 \\
GNB & N/A & 0.48$\pm$0.02 & 0.49$\pm$0.01 & 0.74$\pm$0.03 & 0.59$\pm$0.02 & 0.58$\pm$0.03 & 0.53$\pm$0.02 \\
GB & N/A & 0.68$\pm$0.02 & 0.66$\pm$0.02 & 0.74$\pm$0.02 & 0.70$\pm$0.02 & 0.71$\pm$0.04 & 0.74$\pm$0.03 \\
LGBM & N/A & 0.72$\pm$0.02 & 0.70$\pm$0.02 & 0.79$\pm$0.02 & 0.74$\pm$0.02 & 0.76$\pm$0.04 & 0.79$\pm$0.03 \\
QDA & N/A & 0.53$\pm$0.02 & 0.52$\pm$0.01 & 0.80$\pm$0.03 & 0.63$\pm$0.02 & 0.63$\pm$0.03 & 0.59$\pm$0.02 \\
RF & N/A & 0.72$\pm$0.02 & 0.73$\pm$0.02 & 0.72$\pm$0.03 & 0.72$\pm$0.02 & 0.77$\pm$0.03 & 0.78$\pm$0.02 \\
XGB & N/A & 0.73$\pm$0.02 & 0.70$\pm$0.02 & 0.80$\pm$0.03 & 0.75$\pm$0.02 & 0.77$\pm$0.03 & 0.80$\pm$0.02 \\
\midrule
\multicolumn{8}{l}{\textit{Scale-Sensitive Models}} \\
\midrule
KNN & N & 0.56$\pm$0.02 & 0.57$\pm$0.03 & 0.51$\pm$0.03 & 0.54$\pm$0.02 & 0.57$\pm$0.02 & 0.59$\pm$0.03 \\
KNN & Y & 0.67$\pm$0.02 & 0.66$\pm$0.02 & 0.71$\pm$0.03 & 0.68$\pm$0.02 & 0.67$\pm$0.03 & 0.73$\pm$0.03 \\
MLP & N & 0.51$\pm$0.05 & 0.52$\pm$0.08 & 0.68$\pm$0.24 & 0.56$\pm$0.13 & 0.52$\pm$0.06 & 0.53$\pm$0.08 \\
MLP & Y & 0.72$\pm$0.02 & 0.71$\pm$0.02 & 0.76$\pm$0.04 & 0.73$\pm$0.02 & 0.76$\pm$0.03 & 0.79$\pm$0.03 \\
SVC & N & 0.42$\pm$0.02 & 0.45$\pm$0.01 & 0.68$\pm$0.03 & 0.54$\pm$0.02 & 0.54$\pm$0.02 & 0.46$\pm$0.02 \\
SVC & Y & 0.63$\pm$0.02 & 0.62$\pm$0.02 & 0.69$\pm$0.03 & 0.65$\pm$0.02 & 0.69$\pm$0.03 & 0.70$\pm$0.02 \\
\bottomrule
\end{tabular}%
}

\small
 Models: AB=AdaBoost, DT=Decision Tree, GB=Gradient Boosting,
GNB=Gaussian Naive Bayes, KNN=K-Nearest Neighbors, LGBM=LightGBM, MLP=Multi-Layer Perceptron,
QDA=Quadratic Discriminant Analysis, RF=Random Forest, SVC=Support Vector Classifier (RBF),
XGB=XGBoost. Sc.=Scaling, N/A=not applicable, N=No, Y=Yes.
Bold rows: all metrics $>$ 0.75. Values: mean $\pm$ std over 20 iterations with random resampling.

\end{table*}

\begin{table*}[htb]
\centering
\caption{Model performance: Train on Kepler, Test on Kepler (K/K), {\bf values reported as $\mu\pm\sigma$}}
\label{tab::k_k}
\resizebox{\textwidth}{!}{%
\begin{tabular}{lccccccc}
\toprule
Model & Sc. & Acc & Prec & Rec & F1 & PR-AUC & ROC-AUC \\
\midrule
\multicolumn{8}{l}{\textit{Scale-Invariant Models}} \\
\midrule
{\bf AB} & {\bf N/A} & {\bf 0.85$\pm$0.01} & {\bf 0.81$\pm$0.01} & {\bf 0.92$\pm$0.01} & {\bf 0.86$\pm$0.01} & {\bf 0.89$\pm$0.01} & {\bf 0.92$\pm$0.01} \\
{\bf DT} & {\bf N/A} & {\bf 0.84$\pm$0.01} & {\bf 0.85$\pm$0.01} & {\bf 0.84$\pm$0.01} & {\bf 0.84$\pm$0.01} & {\bf 0.79$\pm$0.01} & {\bf 0.84$\pm$0.01} \\
GNB & N/A & 0.75$\pm$0.01 & 0.67$\pm$0.01 & 0.97$\pm$0.01 & 0.79$\pm$0.01 & 0.82$\pm$0.02 & 0.85$\pm$0.01 \\
{\bf GB} & {\bf N/A} & {\bf 0.88$\pm$0.01} & {\bf 0.85$\pm$0.01} & {\bf 0.93$\pm$0.01} & {\bf 0.89$\pm$0.01} & {\bf 0.94$\pm$0.00} & {\bf 0.95$\pm$0.00} \\
{\bf LGBM} & {\bf N/A} & {\bf 0.90$\pm$0.01} & {\bf 0.88$\pm$0.01} & {\bf 0.92$\pm$0.01} & {\bf 0.90$\pm$0.01} & {\bf 0.95$\pm$0.01} & {\bf 0.96$\pm$0.00} \\
QDA & N/A & 0.78$\pm$0.01 & 0.70$\pm$0.01 & 0.97$\pm$0.01 & 0.81$\pm$0.01 & 0.83$\pm$0.02 & 0.87$\pm$0.01 \\
{\bf RF} & {\bf N/A} & {\bf 0.89$\pm$0.01} & {\bf 0.87$\pm$0.01} & {\bf 0.91$\pm$0.01} & {\bf 0.89$\pm$0.01} & {\bf 0.95$\pm$0.01} & {\bf 0.96$\pm$0.00} \\
{\bf XGB} & {\bf N/A} & {\bf 0.90$\pm$0.01} & {\bf 0.88$\pm$0.01} & {\bf 0.92$\pm$0.01} & {\bf 0.90$\pm$0.01} & {\bf 0.95$\pm$0.00} & {\bf 0.96$\pm$0.00} \\
\midrule
\multicolumn{8}{l}{\textit{Scale-Sensitive Models}} \\
\midrule
KNN & N & 0.74$\pm$0.01 & 0.71$\pm$0.01 & 0.81$\pm$0.02 & 0.76$\pm$0.01 & 0.73$\pm$0.01 & 0.80$\pm$0.01 \\
{\bf KNN} & {\bf Y} & {\bf 0.82$\pm$0.01} & {\bf 0.79$\pm$0.02} & {\bf 0.89$\pm$0.02} & {\bf 0.83$\pm$0.01} & {\bf 0.83$\pm$0.01} & {\bf 0.88$\pm$0.01} \\
MLP & N & 0.75$\pm$0.06 & 0.75$\pm$0.06 & 0.78$\pm$0.22 & 0.74$\pm$0.16 & 0.80$\pm$0.04 & 0.86$\pm$0.02 \\
{\bf MLP} & {\bf Y} & {\bf 0.88$\pm$0.01} & {\bf 0.85$\pm$0.01} & {\bf 0.91$\pm$0.02} & {\bf 0.88$\pm$0.01} & {\bf 0.93$\pm$0.01} & {\bf 0.94$\pm$0.01} \\
SVC & N & 0.64$\pm$0.01 & 0.58$\pm$0.01 & 0.98$\pm$0.01 & 0.73$\pm$0.01 & 0.59$\pm$0.01 & 0.67$\pm$0.01 \\
SVC & Y & 0.78$\pm$0.01 & 0.71$\pm$0.01 & 0.94$\pm$0.01 & 0.81$\pm$0.01 & 0.84$\pm$0.01 & 0.87$\pm$0.01 \\
\bottomrule
\end{tabular}%
}

\small
Models: AB=AdaBoost, DT=Decision Tree, GB=Gradient Boosting,
GNB=Gaussian Naive Bayes, KNN=K-Nearest Neighbors, LGBM=LightGBM, MLP=Multi-Layer Perceptron,
QDA=Quadratic Discriminant Analysis, RF=Random Forest, SVC=Support Vector Classifier (RBF),
XGB=XGBoost. Sc.=Scaling, N/A=not applicable, N=No, Y=Yes.
Bold rows: all metrics $>$ 0.75. Values: mean $\pm$ std over 20 iterations with random resampling.

\end{table*}

\begin{table*}[htb]
\centering
\caption{Model performance: Train on Kepler, Test on TESS+Kepler (K/TK), {\bf values reported as $\mu\pm\sigma$}}
\label{tab::k_tk}
\resizebox{\textwidth}{!}{%
\begin{tabular}{lccccccc}
\toprule
Model & Sc. & Acc & Prec & Rec & F1 & PR-AUC & ROC-AUC \\
\midrule
\multicolumn{8}{l}{\textit{Scale-Invariant Models}} \\
\midrule
AB & N/A & 0.73$\pm$0.01 & 0.70$\pm$0.01 & 0.80$\pm$0.02 & 0.75$\pm$0.01 & 0.78$\pm$0.02 & 0.81$\pm$0.01 \\
DT & N/A & 0.76$\pm$0.01 & 0.75$\pm$0.01 & 0.78$\pm$0.01 & 0.76$\pm$0.01 & 0.69$\pm$0.01 & 0.76$\pm$0.01 \\
GNB & N/A & 0.62$\pm$0.01 & 0.58$\pm$0.01 & 0.85$\pm$0.01 & 0.69$\pm$0.00 & 0.72$\pm$0.01 & 0.72$\pm$0.01 \\
{\bf GB} & {\bf N/A} & {\bf 0.78$\pm$0.01} & {\bf 0.76$\pm$0.01} & {\bf 0.83$\pm$0.01} & {\bf 0.79$\pm$0.01} & {\bf 0.83$\pm$0.01} & {\bf 0.86$\pm$0.01} \\
{\bf LGBM} & {\bf N/A} & {\bf 0.81$\pm$0.01} & {\bf 0.79$\pm$0.01} & {\bf 0.86$\pm$0.01} & {\bf 0.82$\pm$0.01} & {\bf 0.86$\pm$0.01} & {\bf 0.88$\pm$0.00} \\
QDA & N/A & 0.66$\pm$0.01 & 0.61$\pm$0.01 & 0.88$\pm$0.01 & 0.72$\pm$0.00 & 0.75$\pm$0.01 & 0.76$\pm$0.01 \\
{\bf RF} & {\bf N/A} & {\bf 0.81$\pm$0.01} & {\bf 0.80$\pm$0.01} & {\bf 0.81$\pm$0.02} & {\bf 0.81$\pm$0.01} & {\bf 0.87$\pm$0.01} & {\bf 0.88$\pm$0.00} \\
{\bf XGB} & {\bf N/A} & {\bf 0.82$\pm$0.01} & {\bf 0.79$\pm$0.01} & {\bf 0.86$\pm$0.01} & {\bf 0.82$\pm$0.01} & {\bf 0.86$\pm$0.01} & {\bf 0.89$\pm$0.01} \\
\midrule
\multicolumn{8}{l}{\textit{Scale-Sensitive Models}} \\
\midrule
KNN & N & 0.65$\pm$0.01 & 0.65$\pm$0.01 & 0.66$\pm$0.01 & 0.66$\pm$0.01 & 0.65$\pm$0.01 & 0.70$\pm$0.01 \\
KNN & Y & 0.75$\pm$0.01 & 0.73$\pm$0.01 & 0.80$\pm$0.01 & 0.76$\pm$0.01 & 0.76$\pm$0.01 & 0.81$\pm$0.01 \\
MLP & N & 0.63$\pm$0.03 & 0.62$\pm$0.05 & 0.72$\pm$0.20 & 0.64$\pm$0.13 & 0.61$\pm$0.03 & 0.67$\pm$0.03 \\
{\bf MLP} & {\bf Y} & {\bf 0.80$\pm$0.01} & {\bf 0.78$\pm$0.01} & {\bf 0.84$\pm$0.02} & {\bf 0.81$\pm$0.01} & {\bf 0.86$\pm$0.01} & {\bf 0.88$\pm$0.01} \\
SVC & N & 0.53$\pm$0.01 & 0.52$\pm$0.00 & 0.83$\pm$0.01 & 0.64$\pm$0.00 & 0.58$\pm$0.01 & 0.60$\pm$0.01 \\
SVC & Y & 0.71$\pm$0.01 & 0.67$\pm$0.01 & 0.81$\pm$0.01 & 0.73$\pm$0.01 & 0.78$\pm$0.01 & 0.80$\pm$0.01 \\
\bottomrule
\end{tabular}%
}

\small
 Models: AB=AdaBoost, DT=Decision Tree, GB=Gradient Boosting,
GNB=Gaussian Naive Bayes, KNN=K-Nearest Neighbors, LGBM=LightGBM, MLP=Multi-Layer Perceptron,
QDA=Quadratic Discriminant Analysis, RF=Random Forest, SVC=Support Vector Classifier (RBF),
XGB=XGBoost. Sc.=Scaling, N/A=not applicable, N=No, Y=Yes.
Bold rows: all metrics $>$ 0.75. Values: mean $\pm$ std over 20 iterations with random resampling.

\end{table*}

Transitioning to models trained only on the Kepler data (K/T, K/K, and K/TK, in Tables \ref{tab::k_t}, \ref{tab::k_k}, and \ref{tab::k_tk}, respectively), we found an expected result. While testing on Kepler yielded the best metrics of all of our experiments (see Table \ref{tab::k_k}), testing on TESS (see Table \ref{tab::k_t}) was disastrous.  This indicates, of course, that there is a substantial distribution mismatch in our chosen parameters between TESS and Kepler, which we discussed in Section \ref{sec::results}.  However, the lack of a similar result in the opposite case (i.e. T/K) further clarifies the mismatch as primarily being driven by TESS.  { We suggest that this difference is because TESS, with its wide bandpass and broader survey, contains sufficient characteristics of the planet population of the Kepler survey, while the reverse relationship is not nearly as comprehensive.  As such, to create a generalizable model in the case of different distributions, we proceeded to train on the combined TESS and Kepler datasets.}

\subsection{Training on Combined TESS and Kepler Data}
\label{sec::combined_train}

\begin{table*}[htb]
\centering
\caption{Model performance: Train on TESS+Kepler, Test on TESS (TK/T), {\bf values reported as $\mu\pm\sigma$}}
\label{tab::tk_t}
\resizebox{\textwidth}{!}{%
\begin{tabular}{lccccccc}
\toprule
Model & Sc. & Acc & Prec & Rec & F1 & PR-AUC & ROC-AUC \\
\midrule
\multicolumn{8}{l}{\textit{Scale-Invariant Models}} \\
\midrule
AB & N/A & 0.72$\pm$0.02 & 0.70$\pm$0.02 & 0.77$\pm$0.04 & 0.73$\pm$0.02 & 0.75$\pm$0.03 & 0.79$\pm$0.02 \\
DT & N/A & 0.75$\pm$0.02 & 0.74$\pm$0.02 & 0.77$\pm$0.03 & 0.75$\pm$0.02 & 0.68$\pm$0.02 & 0.75$\pm$0.02 \\
GNB & N/A & 0.53$\pm$0.01 & 0.51$\pm$0.00 & 0.98$\pm$0.01 & 0.67$\pm$0.00 & 0.59$\pm$0.02 & 0.61$\pm$0.02 \\
{\bf GB} & {\bf N/A} & {\bf 0.80$\pm$0.02} & {\bf 0.76$\pm$0.02} & {\bf 0.88$\pm$0.02} & {\bf 0.81$\pm$0.02} & {\bf 0.84$\pm$0.03} & {\bf 0.87$\pm$0.02} \\
{\bf LGBM} & {\bf N/A} & {\bf 0.82$\pm$0.01} & {\bf 0.79$\pm$0.02} & {\bf 0.87$\pm$0.02} & {\bf 0.83$\pm$0.01} & {\bf 0.86$\pm$0.02} & {\bf 0.89$\pm$0.01} \\
QDA & N/A & 0.55$\pm$0.01 & 0.53$\pm$0.01 & 0.98$\pm$0.01 & 0.69$\pm$0.00 & 0.70$\pm$0.03 & 0.73$\pm$0.02 \\
{\bf RF} & {\bf N/A} & {\bf 0.81$\pm$0.01} & {\bf 0.79$\pm$0.01} & {\bf 0.87$\pm$0.02} & {\bf 0.82$\pm$0.01} & {\bf 0.85$\pm$0.03} & {\bf 0.89$\pm$0.01} \\
{\bf XGB} & {\bf N/A} & {\bf 0.82$\pm$0.01} & {\bf 0.79$\pm$0.02} & {\bf 0.87$\pm$0.02} & {\bf 0.83$\pm$0.01} & {\bf 0.85$\pm$0.03} & {\bf 0.88$\pm$0.02} \\
\midrule
\multicolumn{8}{l}{\textit{Scale-Sensitive Models}} \\
\midrule
KNN & N & 0.65$\pm$0.02 & 0.64$\pm$0.02 & 0.71$\pm$0.02 & 0.67$\pm$0.01 & 0.65$\pm$0.02 & 0.70$\pm$0.02 \\
KNN & Y & 0.75$\pm$0.01 & 0.71$\pm$0.01 & 0.85$\pm$0.02 & 0.77$\pm$0.01 & 0.76$\pm$0.02 & 0.81$\pm$0.02 \\
MLP & N & 0.56$\pm$0.05 & 0.56$\pm$0.06 & 0.67$\pm$0.21 & 0.59$\pm$0.09 & 0.59$\pm$0.06 & 0.61$\pm$0.08 \\
{\bf MLP} & {\bf Y} & {\bf 0.80$\pm$0.02} & {\bf 0.76$\pm$0.02} & {\bf 0.89$\pm$0.02} & {\bf 0.82$\pm$0.01} & {\bf 0.83$\pm$0.03} & {\bf 0.86$\pm$0.02} \\
SVC & N & 0.57$\pm$0.01 & 0.54$\pm$0.01 & 0.97$\pm$0.01 & 0.69$\pm$0.01 & 0.69$\pm$0.03 & 0.73$\pm$0.02 \\
SVC & Y & 0.72$\pm$0.01 & 0.68$\pm$0.01 & 0.84$\pm$0.02 & 0.75$\pm$0.01 & 0.73$\pm$0.03 & 0.77$\pm$0.02 \\
\bottomrule
\end{tabular}%
}

\small
 Models: AB=AdaBoost, DT=Decision Tree, GB=Gradient Boosting,
GNB=Gaussian Naive Bayes, KNN=K-Nearest Neighbors, LGBM=LightGBM, MLP=Multi-Layer Perceptron,
QDA=Quadratic Discriminant Analysis, RF=Random Forest, SVC=Support Vector Classifier (RBF),
XGB=XGBoost. Sc.=Scaling, N/A=not applicable, N=No, Y=Yes.
Bold rows: all metrics $>$ 0.75. Values: mean $\pm$ std over 20 iterations with random resampling.

\end{table*}

\begin{table*}[htb]
\centering
\caption{Model performance: Train on TESS+Kepler, Test on Kepler (TK/K), {\bf values reported as $\mu\pm\sigma$}}
\label{tab::tk_k}
\resizebox{\textwidth}{!}{%
\begin{tabular}{lccccccc}
\toprule
Model & Sc. & Acc & Prec & Rec & F1 & PR-AUC & ROC-AUC \\
\midrule
\multicolumn{8}{l}{\textit{Scale-Invariant Models}} \\
\midrule
{\bf AB} & {\bf N/A} & {\bf 0.83$\pm$0.02} & {\bf 0.79$\pm$0.02} & {\bf 0.89$\pm$0.03} & {\bf 0.84$\pm$0.02} & {\bf 0.85$\pm$0.02} & {\bf 0.89$\pm$0.01} \\
{\bf DT} & {\bf N/A} & {\bf 0.82$\pm$0.02} & {\bf 0.83$\pm$0.02} & {\bf 0.81$\pm$0.04} & {\bf 0.82$\pm$0.02} & {\bf 0.77$\pm$0.02} & {\bf 0.82$\pm$0.02} \\
GNB & N/A & 0.73$\pm$0.02 & 0.66$\pm$0.01 & 0.98$\pm$0.01 & 0.79$\pm$0.01 & 0.82$\pm$0.03 & 0.85$\pm$0.02 \\
{\bf GB} & {\bf N/A} & {\bf 0.88$\pm$0.01} & {\bf 0.85$\pm$0.01} & {\bf 0.92$\pm$0.01} & {\bf 0.89$\pm$0.01} & {\bf 0.92$\pm$0.01} & {\bf 0.94$\pm$0.01} \\
{\bf LGBM} & {\bf N/A} & {\bf 0.89$\pm$0.01} & {\bf 0.87$\pm$0.02} & {\bf 0.91$\pm$0.02} & {\bf 0.89$\pm$0.01} & {\bf 0.94$\pm$0.01} & {\bf 0.95$\pm$0.01} \\
QDA & N/A & 0.76$\pm$0.02 & 0.69$\pm$0.02 & 0.97$\pm$0.01 & 0.80$\pm$0.01 & 0.84$\pm$0.02 & 0.87$\pm$0.02 \\
{\bf RF} & {\bf N/A} & {\bf 0.89$\pm$0.01} & {\bf 0.87$\pm$0.01} & {\bf 0.90$\pm$0.02} & {\bf 0.89$\pm$0.01} & {\bf 0.93$\pm$0.01} & {\bf 0.95$\pm$0.01} \\
{\bf XGB} & {\bf N/A} & {\bf 0.88$\pm$0.01} & {\bf 0.87$\pm$0.02} & {\bf 0.91$\pm$0.02} & {\bf 0.89$\pm$0.01} & {\bf 0.93$\pm$0.01} & {\bf 0.95$\pm$0.01} \\
\midrule
\multicolumn{8}{l}{\textit{Scale-Sensitive Models}} \\
\midrule
KNN & N & 0.69$\pm$0.02 & 0.67$\pm$0.02 & 0.74$\pm$0.03 & 0.70$\pm$0.02 & 0.68$\pm$0.03 & 0.75$\pm$0.02 \\
{\bf KNN} & {\bf Y} & {\bf 0.82$\pm$0.02} & {\bf 0.78$\pm$0.02} & {\bf 0.88$\pm$0.02} & {\bf 0.83$\pm$0.02} & {\bf 0.83$\pm$0.03} & {\bf 0.88$\pm$0.02} \\
MLP & N & 0.64$\pm$0.11 & 0.60$\pm$0.16 & 0.57$\pm$0.36 & 0.54$\pm$0.29 & 0.71$\pm$0.08 & 0.79$\pm$0.05 \\
{\bf MLP} & {\bf Y} & {\bf 0.88$\pm$0.01} & {\bf 0.87$\pm$0.02} & {\bf 0.89$\pm$0.02} & {\bf 0.88$\pm$0.01} & {\bf 0.91$\pm$0.02} & {\bf 0.93$\pm$0.01} \\
SVC & N & 0.62$\pm$0.01 & 0.57$\pm$0.01 & 1.00$\pm$0.00 & 0.73$\pm$0.01 & 0.62$\pm$0.02 & 0.68$\pm$0.02 \\
SVC & Y & 0.78$\pm$0.02 & 0.72$\pm$0.02 & 0.94$\pm$0.02 & 0.81$\pm$0.01 & 0.83$\pm$0.03 & 0.86$\pm$0.02 \\
\bottomrule
\end{tabular}%
}

\small
 Models: AB=AdaBoost, DT=Decision Tree, GB=Gradient Boosting,
GNB=Gaussian Naive Bayes, KNN=K-Nearest Neighbors, LGBM=LightGBM, MLP=Multi-Layer Perceptron,
QDA=Quadratic Discriminant Analysis, RF=Random Forest, SVC=Support Vector Classifier (RBF),
XGB=XGBoost. Sc.=Scaling, N/A=not applicable, N=No, Y=Yes.
Bold rows: all metrics $>$ 0.75. Values: mean $\pm$ std over 20 iterations with random resampling.

\end{table*}

\begin{table*}[htb]
\centering
\caption{Model performance: Train on TESS+Kepler, Test on TESS+Kepler (TK/TK), {\bf values reported as $\mu\pm\sigma$}}
\label{tab::tk_tk}
\resizebox{\textwidth}{!}{%
\begin{tabular}{lccccccc}
\toprule
Model & Sc. & Acc & Prec & Rec & F1 & PR-AUC & ROC-AUC \\
\midrule
\multicolumn{8}{l}{\textit{Scale-Invariant Models}} \\
\midrule
AB & N/A & 0.77$\pm$0.01 & 0.73$\pm$0.01 & 0.83$\pm$0.02 & 0.78$\pm$0.01 & 0.80$\pm$0.02 & 0.84$\pm$0.01 \\
DT & N/A & 0.78$\pm$0.01 & 0.78$\pm$0.02 & 0.79$\pm$0.02 & 0.79$\pm$0.01 & 0.72$\pm$0.02 & 0.78$\pm$0.01 \\
GNB & N/A & 0.62$\pm$0.01 & 0.57$\pm$0.01 & 0.98$\pm$0.01 & 0.72$\pm$0.00 & 0.69$\pm$0.02 & 0.73$\pm$0.02 \\
{\bf GB} & {\bf N/A} & {\bf 0.83$\pm$0.01} & {\bf 0.80$\pm$0.01} & {\bf 0.90$\pm$0.01} & {\bf 0.84$\pm$0.01} & {\bf 0.88$\pm$0.02} & {\bf 0.91$\pm$0.01} \\
{\bf LGBM} & {\bf N/A} & {\bf 0.85$\pm$0.01} & {\bf 0.82$\pm$0.01} & {\bf 0.89$\pm$0.01} & {\bf 0.86$\pm$0.01} & {\bf 0.90$\pm$0.01} & {\bf 0.92$\pm$0.01} \\
QDA & N/A & 0.65$\pm$0.01 & 0.59$\pm$0.01 & 0.97$\pm$0.01 & 0.74$\pm$0.01 & 0.76$\pm$0.02 & 0.80$\pm$0.01 \\
{\bf RF} & {\bf N/A} & {\bf 0.84$\pm$0.01} & {\bf 0.82$\pm$0.01} & {\bf 0.88$\pm$0.01} & {\bf 0.85$\pm$0.01} & {\bf 0.89$\pm$0.02} & {\bf 0.92$\pm$0.01} \\
{\bf XGB} & {\bf N/A} & {\bf 0.85$\pm$0.01} & {\bf 0.82$\pm$0.01} & {\bf 0.89$\pm$0.01} & {\bf 0.85$\pm$0.01} & {\bf 0.89$\pm$0.02} & {\bf 0.91$\pm$0.01} \\
\midrule
\multicolumn{8}{l}{\textit{Scale-Sensitive Models}} \\
\midrule
KNN & N & 0.67$\pm$0.01 & 0.65$\pm$0.01 & 0.73$\pm$0.02 & 0.69$\pm$0.01 & 0.66$\pm$0.01 & 0.72$\pm$0.01 \\
KNN & Y & 0.78$\pm$0.01 & 0.74$\pm$0.01 & 0.86$\pm$0.01 & 0.80$\pm$0.01 & 0.79$\pm$0.01 & 0.84$\pm$0.01 \\
MLP & N & 0.60$\pm$0.03 & 0.61$\pm$0.04 & 0.63$\pm$0.19 & 0.60$\pm$0.10 & 0.61$\pm$0.04 & 0.66$\pm$0.04 \\
{\bf MLP} & {\bf Y} & {\bf 0.83$\pm$0.01} & {\bf 0.80$\pm$0.02} & {\bf 0.89$\pm$0.02} & {\bf 0.84$\pm$0.01} & {\bf 0.87$\pm$0.02} & {\bf 0.90$\pm$0.01} \\
SVC & N & 0.59$\pm$0.01 & 0.55$\pm$0.01 & 0.99$\pm$0.00 & 0.71$\pm$0.01 & 0.65$\pm$0.02 & 0.70$\pm$0.02 \\
SVC & Y & 0.75$\pm$0.01 & 0.69$\pm$0.01 & 0.88$\pm$0.01 & 0.78$\pm$0.01 & 0.78$\pm$0.02 & 0.82$\pm$0.02 \\
\bottomrule
\end{tabular}%
}

\small
Models: AB=AdaBoost, DT=Decision Tree, GB=Gradient Boosting,
GNB=Gaussian Naive Bayes, KNN=K-Nearest Neighbors, LGBM=LightGBM, MLP=Multi-Layer Perceptron,
QDA=Quadratic Discriminant Analysis, RF=Random Forest, SVC=Support Vector Classifier (RBF),
XGB=XGBoost. Sc.=Scaling, N/A=not applicable, N=No, Y=Yes.
Bold rows: all metrics $>$ 0.75. Values: mean $\pm$ std over 20 iterations with random resampling.

\end{table*}

Models trained on the combined TESS+Kepler data (TK/T, TK/K, and TK/TK, with results shown in Tables \ref{tab::tk_t}, \ref{tab::tk_k}, and \ref{tab::tk_tk}, respectively) have the advantage of training on broader distributions.  As with our other scenarios, the tree-based models and the scaled MLP were all successful by our definition and perform similarly well to the T/T and K/K scenarios, but slightly reduced in most metrics.  We can attribute this reduction in performance to training on the broader distribution of TESS+Kepler, which reduces performance on the tighter distribution of { the individual instruments}, but makes the model more generalizable. 

Testing on TESS+Kepler (TK/TK), however, showed substantially improved results from the T/TK and K/TK scenarios.  This was, again, to be expected.  Training on only TESS or Kepler data cannot possibly create a model with an understanding of such varied parameter distributions between the two.  In recognition of their ability to generalize to the broader distribution, we assess the successful TK/TK models as the strongest.

\subsection{Predictions}
\label{sec::predictions}

All our work described to this point has been demonstrating proof-of-concept regarding the performance of each of the models in identifying confirmed exoplanets and confirmed false positives.  { Here, using the best trained models that we described in Section \ref{sec::combined_train}, we show our predictions of the TESS and Kepler unresolved planet candidates.}

The results of the top three models (LGBM, XGB, and RF) are rather similar.  These are all tree-based models, indicating the strength of that methodology for this particular task.  {The scaled MLP model, with an entirely different methodology, follows closely behind these models in overall performance.  Each of them, however, certainly has strengths and weaknesses that would be cumbersome to diagnose.  As such, we combined them into an {\it ensemble} of three models (MLP, XGB, and RF), where MLP was substituted for LGBM to represent a greater diversity of methodologies and produce more robust results.  As an ensemble, the combination of disparate models was used to create an average prediction robust to disparate gaps in any individual model.}

\begin{figure*}[htp]
    \centering
    \includegraphics[width=\textwidth]{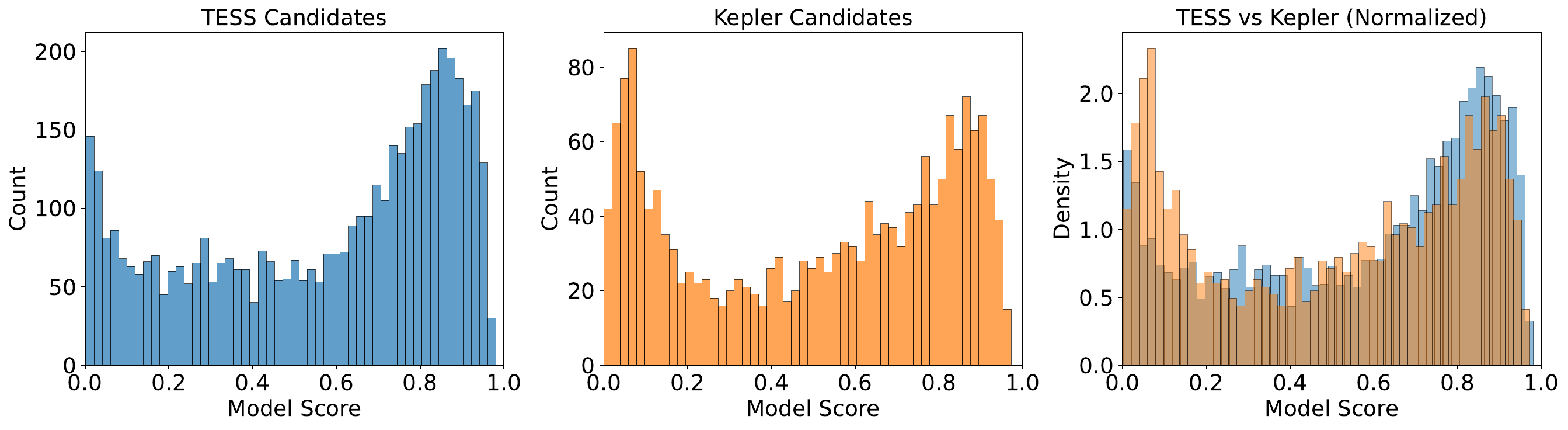}
    \caption{Distribution of ensemble model prediction scores for planet candidates from TESS and Kepler. ({\em left}) TESS candidates (n=4701).  ({\em center}) Kepler candidates (n=1874). ({\em right}) Normalized overlay comparison. Each candidate's overall score is the mean of 60 predictions (3 models × 20 iterations) from {MLP, RF, and XGB classifiers} trained on balanced confirmed planets and false positives from both surveys.}
    \label{fig::prediction_hist}
\end{figure*}

\begin{figure*}[htp]
    \centering
    \includegraphics[width=\textwidth]{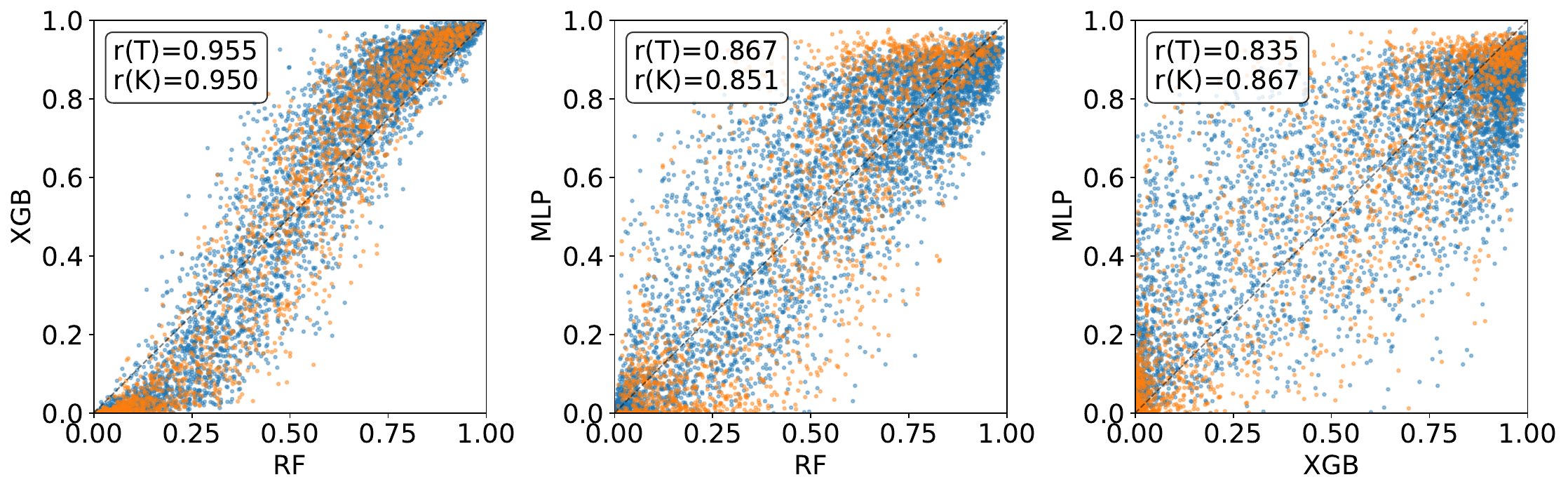}
    \caption{Pairwise comparison of model predictions for TESS (blue) and Kepler (orange) planet candidates. {({\em left}) RF vs XGB, ({\em middle}) RF vs MLP, and ({\em right}) XGB vs MLP.}  Each panel shows the predicted probability from one model versus another. Dashed diagonal line indicates perfect agreement. Correlation coefficients for TESS and Kepler are given by r(T) and r(K) on each plot, respectively.}
    \label{fig::prediction_corr}
\end{figure*}

\begin{figure*}[]
    \centering
    \includegraphics[width=\textwidth]{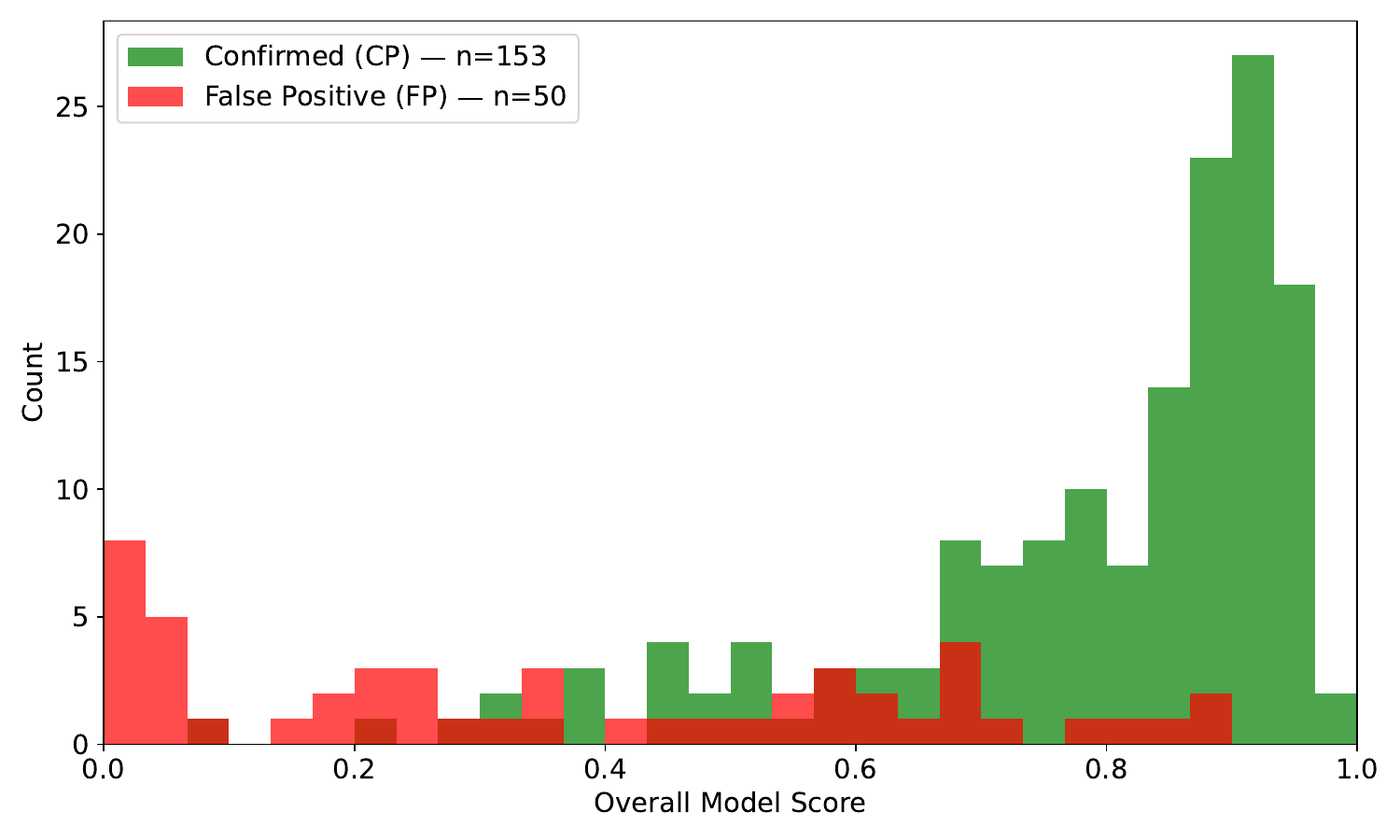}
    \caption{Ensemble model scores for 203 TOIs resolved as either CP (green) or FP (red) between the dates of the data we used for our analysis (August 30, 2025 and April 28, 2026). See Section \ref{sec::predictions} for details.}
    \label{fig::resolved_toi}
\end{figure*}

Using the random resampling that we described in Section \ref{sec::methodology}, we made 20 predictions from each of our models against the entire set of candidates for both TESS and Kepler, combining sixty distinct scores into the ensemble average score.  We report the top candidates for TESS and Kepler along with the scores for each model in Tables \ref{tab::top_50_tess} and \ref{tab::top_50_kepler}, respectively.  {We also provide, as {digital} supplements to this paper, the complete lists evaluating all TESS and Kepler candidates with our ensemble model.}  We show the distributions of the model predictions in Figure \ref{fig::prediction_hist} for TESS (left), Kepler (center) and both combined (right).  The distributions are as expected for a well-trained model, with the concentration of scores near unity and zero indicating strong positive and negative predictions with ambiguous candidates in between.  

{We also show the correlations of the average prediction for each model in Figure \ref{fig::prediction_corr}.  RF vs XGB is shown on the left, RF vs MLP in the center, and XGB vs MLP on the right.  We had originally used both XGB and LGBM as part of our ensemble model, but we noticed that these models were highly correlated ($r\approx0.99$) and so we replaced LGBM with the scaled MLP (which scored similarly to LGBM in the TK-trained cases) in order to ensure a greater diversity of opinion among the models.  The middle and right panels of Figure \ref{fig::prediction_corr} indeed show the diversity introduced by the MLP.}

\begin{table*}[htb]
\centering
\caption{Top 50 TESS Planet Candidates Ranked by Ensemble Model Score}
\label{tab::top_50_tess}
\begin{tabular}{rllcccc}
\toprule
Rank & TIC ID & TOI & Overall & RF & XGB & MLP \\
\midrule
1 & 323132914 & 1057.01 & 0.981$\pm$0.020 & 0.989$\pm$0.011 & 0.996$\pm$0.003 & 0.957$\pm$0.013 \\
2 & {\bf 468983280} & {\bf 5489.02\textsuperscript{a}} & 0.978$\pm$0.020 & 0.983$\pm$0.012 & 0.996$\pm$0.003 & 0.955$\pm$0.011 \\
3 & 408416746 & 4324.01 & 0.977$\pm$0.020 & 0.970$\pm$0.027 & 0.992$\pm$0.006 & 0.969$\pm$0.008 \\
4 & 441642457 & 2073.01 & 0.975$\pm$0.020 & 0.978$\pm$0.018 & 0.990$\pm$0.010 & 0.957$\pm$0.015 \\
5 & 232650365 & 1746.02 & 0.974$\pm$0.020 & 0.970$\pm$0.018 & 0.996$\pm$0.002 & 0.957$\pm$0.010 \\
6 & 219195044 & 714.02 & 0.973$\pm$0.018 & 0.983$\pm$0.015 & 0.981$\pm$0.014 & 0.954$\pm$0.009 \\
7 & 219229644 & 870.01 & 0.972$\pm$0.018 & 0.974$\pm$0.014 & 0.986$\pm$0.010 & 0.955$\pm$0.013 \\
8 & 441798995 & 2269.03 & 0.972$\pm$0.027 & 0.965$\pm$0.033 & 0.982$\pm$0.029 & 0.968$\pm$0.008 \\
9 & 22903436 & 7032.01 & 0.971$\pm$0.025 & 0.984$\pm$0.015 & 0.987$\pm$0.007 & 0.941$\pm$0.015 \\
10 & 280803800 & 4410.01 & 0.970$\pm$0.024 & 0.982$\pm$0.014 & 0.988$\pm$0.006 & 0.941$\pm$0.012 \\
11 & 271596225 & 797.02 & 0.969$\pm$0.024 & 0.963$\pm$0.026 & 0.989$\pm$0.017 & 0.955$\pm$0.009 \\
12 & {\bf 237099296} & {\bf 1750.01\textsuperscript{b}} & 0.967$\pm$0.033 & 0.990$\pm$0.010 & 0.989$\pm$0.006 & 0.923$\pm$0.013 \\
13 & 92439980 & 5813.01 & 0.967$\pm$0.036 & 0.990$\pm$0.011 & 0.990$\pm$0.007 & 0.921$\pm$0.022 \\
14 & 182078202 & 4302.01 & 0.967$\pm$0.028 & 0.981$\pm$0.019 & 0.983$\pm$0.016 & 0.935$\pm$0.016 \\
15 & 321669174 & 2081.02 & 0.966$\pm$0.025 & 0.958$\pm$0.026 & 0.992$\pm$0.006 & 0.950$\pm$0.010 \\
16 & 335540507 & 5526.01 & 0.966$\pm$0.029 & 0.976$\pm$0.017 & 0.989$\pm$0.007 & 0.933$\pm$0.022 \\
17 & 155012162 & 5548.01 & 0.966$\pm$0.037 & 0.988$\pm$0.015 & 0.993$\pm$0.004 & 0.918$\pm$0.019 \\
18 & 326386495 & 4424.01 & 0.966$\pm$0.028 & 0.978$\pm$0.015 & 0.990$\pm$0.006 & 0.930$\pm$0.013 \\
19 & {\bf 147821208} & {\bf 6393.01\textsuperscript{b}} & 0.966$\pm$0.032 & 0.988$\pm$0.011 & 0.986$\pm$0.007 & 0.923$\pm$0.011 \\
20 & 258776466 & 7391.01 & 0.965$\pm$0.030 & 0.981$\pm$0.021 & 0.985$\pm$0.011 & 0.929$\pm$0.015 \\
21 & 237920046 & 873.01 & 0.964$\pm$0.032 & 0.988$\pm$0.013 & 0.980$\pm$0.016 & 0.925$\pm$0.014 \\
22 & 459970307 & 1154.01 & 0.964$\pm$0.029 & 0.972$\pm$0.026 & 0.987$\pm$0.011 & 0.934$\pm$0.017 \\
23 & {\bf 219698776} & {\bf 1243.01\textsuperscript{c}} & 0.964$\pm$0.021 & 0.949$\pm$0.022 & 0.983$\pm$0.013 & 0.960$\pm$0.010 \\
24 & 236934937 & 2291.01 & 0.964$\pm$0.025 & 0.963$\pm$0.029 & 0.961$\pm$0.031 & 0.967$\pm$0.008 \\
25 & 173002823 & 5710.01 & 0.963$\pm$0.028 & 0.972$\pm$0.018 & 0.988$\pm$0.009 & 0.931$\pm$0.015 \\
26 & 23961340 & 7062.01 & 0.963$\pm$0.026 & 0.963$\pm$0.021 & 0.989$\pm$0.008 & 0.936$\pm$0.012 \\
27 & 318180448 & 5033.01 & 0.962$\pm$0.026 & 0.943$\pm$0.030 & 0.985$\pm$0.012 & 0.960$\pm$0.010 \\
28 & 161477033 & 553.03 & 0.962$\pm$0.029 & 0.966$\pm$0.024 & 0.988$\pm$0.011 & 0.933$\pm$0.016 \\
29 & 428988981 & 2144.01 & 0.962$\pm$0.029 & 0.949$\pm$0.035 & 0.988$\pm$0.009 & 0.949$\pm$0.013 \\
30 & 167342439 & 707.01 & 0.961$\pm$0.021 & 0.954$\pm$0.024 & 0.976$\pm$0.016 & 0.953$\pm$0.013 \\
31 & 154940895 & 4572.01 & 0.960$\pm$0.021 & 0.965$\pm$0.020 & 0.964$\pm$0.028 & 0.951$\pm$0.009 \\
32 & 393546540 & 5696.02 & 0.960$\pm$0.027 & 0.962$\pm$0.025 & 0.984$\pm$0.010 & 0.934$\pm$0.014 \\
33 & 420529106 & 6558.01 & 0.960$\pm$0.028 & 0.986$\pm$0.011 & 0.967$\pm$0.019 & 0.927$\pm$0.010 \\
34 & 354518617 & 2540.02 & 0.960$\pm$0.027 & 0.961$\pm$0.018 & 0.986$\pm$0.012 & 0.933$\pm$0.018 \\
35 & 224298134 & 2079.02 & 0.960$\pm$0.031 & 0.950$\pm$0.043 & 0.963$\pm$0.027 & 0.966$\pm$0.009 \\
36 & 456260074 & 6249.02 & 0.959$\pm$0.025 & 0.957$\pm$0.028 & 0.981$\pm$0.011 & 0.940$\pm$0.014 \\
37 & 345450130 & 5525.01 & 0.959$\pm$0.038 & 0.941$\pm$0.043 & 0.997$\pm$0.001 & 0.940$\pm$0.014 \\
38 & 230387153 & 2086.01 & 0.959$\pm$0.027 & 0.963$\pm$0.020 & 0.983$\pm$0.017 & 0.932$\pm$0.017 \\
39 & 149788158 & 727.01 & 0.959$\pm$0.029 & 0.936$\pm$0.031 & 0.983$\pm$0.019 & 0.959$\pm$0.009 \\
40 & 219195044 & 714.01 & 0.959$\pm$0.026 & 0.957$\pm$0.023 & 0.983$\pm$0.013 & 0.936$\pm$0.014 \\
41 & 307849973 & 4567.02 & 0.958$\pm$0.033 & 0.975$\pm$0.017 & 0.979$\pm$0.021 & 0.921$\pm$0.022 \\
42 & 18019350 & 5397.01 & 0.958$\pm$0.033 & 0.981$\pm$0.019 & 0.975$\pm$0.019 & 0.919$\pm$0.015 \\
43 & 7422496 & 4189.01 & 0.958$\pm$0.023 & 0.952$\pm$0.023 & 0.972$\pm$0.022 & 0.950$\pm$0.016 \\
44 & 38584799 & 6633.01 & 0.958$\pm$0.033 & 0.982$\pm$0.014 & 0.973$\pm$0.021 & 0.919$\pm$0.010 \\
45 & 234345288 & 213.01 & 0.958$\pm$0.031 & 0.964$\pm$0.029 & 0.981$\pm$0.018 & 0.929$\pm$0.021 \\
46 & 271596225 & 797.03 & 0.958$\pm$0.033 & 0.955$\pm$0.023 & 0.995$\pm$0.004 & 0.924$\pm$0.016 \\
47 & 467158766 & 5714.01 & 0.958$\pm$0.040 & 0.987$\pm$0.011 & 0.982$\pm$0.009 & 0.904$\pm$0.017 \\
48 & 232624234 & 5611.01 & 0.958$\pm$0.035 & 0.982$\pm$0.009 & 0.979$\pm$0.014 & 0.912$\pm$0.014 \\
49 & 147923561 & 5715.01 & 0.957$\pm$0.025 & 0.961$\pm$0.024 & 0.978$\pm$0.014 & 0.934$\pm$0.013 \\
50 & 300710077 & 789.01 & 0.957$\pm$0.030 & 0.962$\pm$0.023 & 0.969$\pm$0.039 & 0.940$\pm$0.015 \\
\bottomrule
\end{tabular}

\small
[a] Confirmed by \citet{2026AJ....171...99G}.
[b] Confirmed by \citet{2026MNRAS.tmp..514L}.
[c] Confirmed by \citet{2026A&A...707A.106P}.

\end{table*}

\begin{table*}[htb]
\centering
\caption{Top 50 Kepler Planet Candidates Ranked by Ensemble Model Score}
\label{tab::top_50_kepler}
{%
\begin{tabular}{rllcccc}
\toprule
Rank & KIC ID & KOI & Overall & RF & XGB & MLP \\
\midrule
1 & 5087223 & K06513.01 & 0.973$\pm$0.019 & 0.968$\pm$0.021 & 0.990$\pm$0.007 & 0.959$\pm$0.008 \\
2 & 6776555 & K05327.01 & 0.970$\pm$0.024 & 0.968$\pm$0.026 & 0.982$\pm$0.026 & 0.959$\pm$0.009 \\
3 & 3765917 & K04526.01 & 0.969$\pm$0.020 & 0.966$\pm$0.019 & 0.982$\pm$0.015 & 0.958$\pm$0.016 \\
4 & 12208631 & K02449.01 & 0.967$\pm$0.028 & 0.952$\pm$0.037 & 0.991$\pm$0.005 & 0.956$\pm$0.012 \\
5 & 4172746 & K02531.01 & 0.965$\pm$0.029 & 0.963$\pm$0.030 & 0.992$\pm$0.004 & 0.940$\pm$0.014 \\
6 & 11905398 & K05940.01 & 0.963$\pm$0.033 & 0.979$\pm$0.019 & 0.987$\pm$0.015 & 0.923$\pm$0.017 \\
7 & 4242147 & K01934.01 & 0.962$\pm$0.033 & 0.972$\pm$0.020 & 0.990$\pm$0.009 & 0.924$\pm$0.022 \\
8 & 6679295 & K02862.01 & 0.962$\pm$0.027 & 0.952$\pm$0.035 & 0.967$\pm$0.025 & 0.966$\pm$0.012 \\
9 & 5511659 & K04541.02 & 0.960$\pm$0.027 & 0.959$\pm$0.030 & 0.984$\pm$0.010 & 0.938$\pm$0.012 \\
10 & 9580992 & K02217.01 & 0.960$\pm$0.029 & 0.956$\pm$0.028 & 0.989$\pm$0.007 & 0.935$\pm$0.013 \\
11 & 9100484 & K01399.01 & 0.959$\pm$0.036 & 0.980$\pm$0.014 & 0.986$\pm$0.011 & 0.912$\pm$0.012 \\
12 & 5956656 & K01053.02 & 0.959$\pm$0.026 & 0.952$\pm$0.028 & 0.970$\pm$0.030 & 0.954$\pm$0.016 \\
13 & 6665064 & K06751.01 & 0.957$\pm$0.029 & 0.971$\pm$0.026 & 0.969$\pm$0.023 & 0.931$\pm$0.019 \\
14 & 3245969 & K01101.02 & 0.956$\pm$0.029 & 0.951$\pm$0.031 & 0.985$\pm$0.010 & 0.932$\pm$0.011 \\
15 & 10141900 & K01082.04 & 0.954$\pm$0.028 & 0.953$\pm$0.032 & 0.976$\pm$0.017 & 0.934$\pm$0.012 \\
16 & 8313667 & K01145.01 & 0.953$\pm$0.034 & 0.958$\pm$0.036 & 0.955$\pm$0.043 & 0.946$\pm$0.017 \\
17 & 11361283 & K02551.01 & 0.951$\pm$0.032 & 0.951$\pm$0.029 & 0.981$\pm$0.012 & 0.921$\pm$0.015 \\
18 & 9349757 & K03348.02 & 0.951$\pm$0.032 & 0.933$\pm$0.035 & 0.979$\pm$0.020 & 0.940$\pm$0.012 \\
19 & 4676964 & K03069.01 & 0.950$\pm$0.049 & 0.925$\pm$0.046 & 0.957$\pm$0.063 & 0.966$\pm$0.014 \\
20 & 10813078 & K05831.01 & 0.949$\pm$0.037 & 0.967$\pm$0.029 & 0.974$\pm$0.018 & 0.905$\pm$0.013 \\
21 & 5511659 & K04541.01 & 0.948$\pm$0.033 & 0.950$\pm$0.039 & 0.974$\pm$0.010 & 0.920$\pm$0.014 \\
22 & 3539231 & K04626.01 & 0.947$\pm$0.045 & 0.904$\pm$0.048 & 0.977$\pm$0.023 & 0.958$\pm$0.017 \\
23 & 5553959 & K03377.01 & 0.946$\pm$0.041 & 0.904$\pm$0.037 & 0.977$\pm$0.026 & 0.957$\pm$0.011 \\
24 & 7772914 & K03339.01 & 0.946$\pm$0.041 & 0.944$\pm$0.043 & 0.981$\pm$0.015 & 0.913$\pm$0.025 \\
25 & 3219643 & K03072.01 & 0.945$\pm$0.036 & 0.957$\pm$0.033 & 0.966$\pm$0.022 & 0.913$\pm$0.026 \\
26 & 9474756 & K03495.01 & 0.945$\pm$0.033 & 0.941$\pm$0.029 & 0.969$\pm$0.029 & 0.924$\pm$0.024 \\
27 & 9427402 & K01397.01 & 0.944$\pm$0.036 & 0.935$\pm$0.037 & 0.937$\pm$0.044 & 0.960$\pm$0.008 \\
28 & 2861126 & K04957.02 & 0.944$\pm$0.045 & 0.906$\pm$0.048 & 0.988$\pm$0.014 & 0.937$\pm$0.015 \\
29 & 4274548 & K01331.01 & 0.942$\pm$0.027 & 0.941$\pm$0.026 & 0.954$\pm$0.035 & 0.932$\pm$0.011 \\
30 & 6837146 & K01362.01 & 0.942$\pm$0.034 & 0.936$\pm$0.040 & 0.966$\pm$0.028 & 0.924$\pm$0.015 \\
31 & 6665860 & K04840.01 & 0.942$\pm$0.038 & 0.929$\pm$0.034 & 0.981$\pm$0.016 & 0.915$\pm$0.021 \\
32 & 4349442 & K01803.01 & 0.941$\pm$0.040 & 0.955$\pm$0.027 & 0.975$\pm$0.017 & 0.894$\pm$0.017 \\
33 & 3249334 & K06318.01 & 0.941$\pm$0.041 & 0.928$\pm$0.050 & 0.969$\pm$0.033 & 0.927$\pm$0.012 \\
34 & 4670217 & K02560.01 & 0.941$\pm$0.047 & 0.936$\pm$0.055 & 0.978$\pm$0.026 & 0.908$\pm$0.021 \\
35 & 8949925 & K02972.02 & 0.941$\pm$0.039 & 0.912$\pm$0.040 & 0.964$\pm$0.033 & 0.945$\pm$0.021 \\
36 & 2973386 & K03034.01 & 0.940$\pm$0.049 & 0.911$\pm$0.053 & 0.943$\pm$0.051 & 0.967$\pm$0.016 \\
37 & 9966219 & K03401.01 & 0.940$\pm$0.036 & 0.923$\pm$0.033 & 0.977$\pm$0.016 & 0.920$\pm$0.022 \\
38 & 9896018 & K02579.03 & 0.939$\pm$0.037 & 0.927$\pm$0.036 & 0.976$\pm$0.023 & 0.916$\pm$0.017 \\
39 & 7835312 & K06164.01 & 0.939$\pm$0.056 & 0.907$\pm$0.062 & 0.947$\pm$0.060 & 0.964$\pm$0.013 \\
40 & 5542466 & K01590.03 & 0.939$\pm$0.040 & 0.925$\pm$0.046 & 0.980$\pm$0.011 & 0.912$\pm$0.012 \\
41 & 4939265 & K06475.01 & 0.938$\pm$0.038 & 0.926$\pm$0.044 & 0.974$\pm$0.015 & 0.914$\pm$0.015 \\
42 & 4384909 & K07693.01 & 0.938$\pm$0.032 & 0.934$\pm$0.039 & 0.963$\pm$0.021 & 0.919$\pm$0.014 \\
43 & 11520114 & K05909.01 & 0.938$\pm$0.044 & 0.935$\pm$0.061 & 0.958$\pm$0.037 & 0.921$\pm$0.009 \\
44 & 8007174 & K05458.01 & 0.937$\pm$0.038 & 0.941$\pm$0.036 & 0.956$\pm$0.040 & 0.916$\pm$0.024 \\
45 & 5027859 & K05117.01 & 0.937$\pm$0.038 & 0.936$\pm$0.035 & 0.960$\pm$0.038 & 0.915$\pm$0.026 \\
46 & 11657614 & K03370.01 & 0.937$\pm$0.041 & 0.917$\pm$0.046 & 0.965$\pm$0.033 & 0.928$\pm$0.025 \\
47 & 2715695 & K03077.02 & 0.936$\pm$0.046 & 0.962$\pm$0.027 & 0.961$\pm$0.042 & 0.886$\pm$0.012 \\
48 & 8396405 & K01733.01 & 0.936$\pm$0.039 & 0.907$\pm$0.035 & 0.972$\pm$0.025 & 0.928$\pm$0.022 \\
49 & 4270799 & K04810.01 & 0.935$\pm$0.049 & 0.901$\pm$0.055 & 0.979$\pm$0.027 & 0.927$\pm$0.019 \\
50 & 7868967 & K06925.01 & 0.935$\pm$0.064 & 0.910$\pm$0.056 & 0.933$\pm$0.089 & 0.963$\pm$0.008 \\
\bottomrule
\end{tabular}%
}
\end{table*}

\begin{deluxetable*}{cccc}
\tablewidth{0pt}
\tablecaption{Model accuracy at different score thresholds for newly resolved TOIs. \label{tab:thresholds}}
\tablecolumns{4}
\tablehead{
\colhead{Threshold} & \colhead{CP Correct} & \colhead{FP Correct} & \colhead{Overall Accuracy}
}
\startdata
0.05 & 153/153 (100.0\%) & 12/50 (24.0\%) & 165/203 (81.3\%) \\
0.10 & 152/153 (99.3\%) & 14/50 (28.0\%) & 166/203 (81.8\%) \\
0.15 & 152/153 (99.3\%) & 15/50 (30.0\%) & 167/203 (82.3\%) \\
0.20 & 152/153 (99.3\%) & 17/50 (34.0\%) & 169/203 (83.3\%) \\
0.25 & 151/153 (98.7\%) & 21/50 (42.0\%) & 172/203 (84.7\%) \\
0.30 & 150/153 (98.0\%) & 24/50 (48.0\%) & 174/203 (85.7\%) \\
0.35 & 147/153 (96.1\%) & 27/50 (54.0\%) & 174/203 (85.7\%) \\
0.40 & 144/153 (94.1\%) & 28/50 (56.0\%) & 172/203 (84.7\%) \\
0.45 & 140/153 (91.5\%) & 29/50 (58.0\%) & 169/203 (83.3\%) \\
0.50 & 138/153 (90.2\%) & 31/50 (62.0\%) & 169/203 (83.3\%) \\
0.55 & 133/153 (86.9\%) & 34/50 (68.0\%) & 167/203 (82.3\%) \\
0.60 & 130/153 (85.0\%) & 37/50 (74.0\%) & 167/203 (82.3\%) \\
0.65 & 126/153 (82.4\%) & 39/50 (78.0\%) & 165/203 (81.3\%) \\
0.70 & 116/153 (75.8\%) & 44/50 (88.0\%) & 160/203 (78.8\%) \\
0.75 & 108/153 (70.6\%) & 45/50 (90.0\%) & 153/203 (75.4\%) \\
0.80 & 91/153 (59.5\%) & 46/50 (92.0\%) & 137/203 (67.5\%) \\
0.85 & 77/153 (50.3\%) & 48/50 (96.0\%) & 125/203 (61.6\%) \\
0.90 & 47/153 (30.7\%) & 50/50 (100.0\%) & 97/203 (47.8\%) \\
0.95 & 6/153 (3.9\%) & 50/50 (100.0\%) & 56/203 (27.6\%) 
\enddata
\end{deluxetable*}

{After our analysis was completed, as of 28 April 2026, we examined the ``Confirmed Planets'' and ``TESS Confirmed Planets'' tables from the NASA Exoplanet Archive and found that 203 of the planet candidates as of the date of the data we used for our analysis (30 August 2025) had since been resolved.  We show a histogram of the ensemble model scores of each of these in Figure \ref{fig::resolved_toi}.  The figure clearly demonstrates that the distribution of the model scores of the newly confirmed planets (green) is heavily weighted toward high scores, while the opposite is true for the resolved false positives (red). We provide the accuracy of our classifications at score thresholds from 0-1 in increments of 0.05 in Table \ref{tab:thresholds}.  It can be seen this table that we can expect all candidates with ensemble model scores $>0.90$ to be confirmed.  Indeed, looking at our top 50 TESS candidates from Table \ref{tab::top_50_tess}, our number 2 ranked candidate, TOI 5489.02 (ensemble score 0.986), was confirmed as a planet by \citet{2026AJ....171...99G}, our number 12 ranked candidate, TOI 1750.01 (ensemble score 0.967) was confirmed by  \citet{2026MNRAS.tmp..514L}, our number 19 ranked candidate, TOI 6393.01 (ensemble score 0.966), was confirmed by \citet{2026MNRAS.tmp..514L}, and our number 23 ranked candidate, TOI 1243.01 (ensemble score 0.964), was confirmed by \citet{2026A&A...707A.106P}.  Even when reducing the threshold to $>0.80$, we can expect more than 90\% of the candidates to be confirmed.  Based on these results, we assert that our ensemble model method, combining the scores of 60 separate, high-performing models, is highly robust.  We consider additional effort to confirm our top candidates as very likely to be productive.}

\section{Discussion}
\label{sec::discussion}

We are not nearly the first researchers to attempt to develop automated tools for exoplanet candidate vetting. Since the first Kepler data, such methods have been rapidly gaining in sophistication and popularity.   \citet{2012ApJ...761....6M}, \citet{2015ApJ...806....6M},  \citet{2018ApJS..235...38T}, \citet{2019AJ....158...25Y}, \citet{2019AJ....157..124K}, \citet{2021AJ....161...24G}, \citet{2022ApJ...926..120V}, \citet{2024AJ....167..233H}, \citet{Kunimoto_2025}, \citet{2025arXiv250917645H}, and \citet{2025arXiv251108768D}, among others, have made substantial progress to this end, many using machine learning methods in their approaches.  { We note specifically the ExoMiner++ method of \citet{2025AJ....170..287V}, which uses Kepler data to enhance model training for application to TESS and examines scenarios similar to our T/T, K/K, and TK/T with their model.} 

The motivation for the breadth of research in this area is clear: More data and competition for the same resources requires prioritization. The Nancy Grace Roman Space Telescope (Roman) is expected to yield between 60,000 and 200,000 transiting exoplanets \citep{2023ApJS..269....5W}.  {As of 28 April 2026, there are 6,273 confirmed exoplanets\footnote{Per NASA's exoplanet archive, \url{https://exoplanetarchive.ipac.caltech.edu/}}.}  The order of magnitude increase in exoplanet candidates will quickly become intractable for the same quality of examination that they have received in the past, especially considering a relatively constant quantity of available resources to confirm them and scientists to examine them.  Among the most important considerations in selecting a limited number of candidates for follow-up observation {and/or analysis} is the likelihood of the follow-up bearing fruit. 

{ While the general trend of the literature shows more complicated models with more inputs and more processing requirements, our use of a small set of post-processed parameters makes our method's application universal to any existing or future instrument and robust to the biases thereof.  Emphasizing the simplicity, our model can be run on a common laptop computer and produce results within seconds.}  By training our model together on the combined TESS and Kepler post-processed parameters, we have expanded the distribution of real data that the model can interpret and enhanced the model's ability to generalize.  This is critical for application of the model beyond TESS and Kepler.  As such, our model provides an excellent starting point for exoplanet candidate {prioritization} with Roman data.  

{ We of course expect some initial difficulty in adapting our model to Roman planet candidates without additional training, as the photometry of the instrument and the different distribution of planets to which it is sensitive (see, e.g., Figure 9 of \citealt{2019ApJS..241....3P}) will cause the model to struggle without having been trained on data within these distributions, as with our K/T scenario.  We expect, however, that with only on the order of a few hundred training samples of CPs and FPs from Roman, our generalizable model will nicely adapt to Roman data and could be used successfully as a tool for planet candidate prioritization.}

Expanding on the concept of the generalizability, we revisit the distinct TESS and Kepler distributions for each of our chosen parameters.  The KS tests we performed (results shown in Figure \ref{fig::dist}) very clearly concluded what a visual examination of the distributions easily confirms: The distributions are very different.  This is, of course, attributable to differences in the instruments.  TESS extended the distribution of exoplanet characteristics to those orbiting cooler M dwarf stars due to the design of its bandpass.  Kepler, again by design, fixed its field of view on a section of the sky containing distant stars, while TESS has examined nearly the entire sky, capturing entirely different populations.  Roman will also expand our understanding of exoplanet populations through a deep examination the galactic bulge \citep{2023ApJS..269....5W}, which TESS was hardly able to sufficiently resolve with its 21\arcsec pixels.  Post-processing can mitigate instrumental artifacts, creating a universal set of parameters comparable between instruments, but it is constrained by the limits of the instrument itself.

It is clear, both from our results in this paper as well as a common-sense assessment, that the post-processed parameters from a single instrument are best suited to train models to be used with that particular instrument.  Testing outside the distribution, as we saw with our T/K, K/T, T/TK, and K/TK scenarios, demonstrates limited success.  It is with this understanding that we pursue applicability not only to the existing data, but also to future data.  Our ensemble model enhances the strengths and mitigates the weaknesses of instrument-specific models.  

{We encourage the community to consider a closer evaluation of our top TESS candidates, as they are almost certainly confirmable as exoplanets, as demonstrated by the the performance of our model against the 203 recently resolved candidates shown in Figure \ref{fig::resolved_toi}.}  Although recent work on Kepler candidates is understandably less prevalent, we also encourage evaluation of our top Kepler candidates as our model is no less powerful in these predictions. It can easily be argued that, as the data has been available for much longer, these candidates have been thoroughly analyzed and those with a higher likelihood of being confirmed have already been evaluated.  However, our model scores suggest that there still remains a substantial number of planet candidates in the Kepler data that are likely to be confirmed.


\section{Summary}
\label{sec::summary}

We have developed a novel ML approach to predict the likelihood of exoplanet candidate confirmation both the Kepler and TESS post-processed databases. We used eleven different types of ML models with scaled and unscaled data (where appropriate) in a Monte Carlo style random resampling, {concluding that the RF, MLP, and XGB models performed the best for our task while providing for robustness together in an ensemble.}  In all cases, we used the default models available through the python {\sc scikit-learn}, {\sc lightgbm}, and {\sc xgboost} libraries. 

While the component models that we used to make our predictions are not in themselves novel, { our process and the simplicity thereof is the novel aspect of this research.}  We pursued a model capable of generalization across the two most prolific instruments for planet transit detections using post-processed parameters. We examined the distributions of our selected parameters and found them to be substantially different between the two instruments. Using our selected parameters, we evaluated their predictive capability in all nine possible train/test combinations of the data, demonstrating that, while the models trained on the distributions of one instrument struggle to predict outcomes for the other, training on their data jointly yielded a stronger, more generalizable model.  After demonstrating successful performance on both TESS and Kepler datasets, we proceeded to make predictions using planetary candidates.

To make predictions on the unconfirmed planetary candidates for both TESS and Kepler, we used the three indicated models together in an ensemble, averaging the scores of twenty iterations each of random resampling, ensuring that our predictions were robust.  We provided the top fifty candidates from our ensemble model for both TESS and Kepler in this paper as Tables \ref{tab::top_50_tess} and \ref{tab::top_50_kepler}, respectively, while also providing the complete set of prediction scores for all planet candidates as digital supplements to this paper.  

{ In an examination of all 203 TESS planet candidates that were resolved since the date of the data we used for our analysis (153 confirmed planets and 50 false positives), we found that {\em all} of our model predictions were correct in cases where the predicted score was $>=0.90$ (47 confirmed planets) or $<=0.05$ (12 false positives). Further softening the threshold to $>=0.80$ produced 91 confirmed planets and 4 false positives, while a threshold of $<=0.20$ produced 1 confirmed planet and 17 false positives.  These results demonstrate a strong correlation between our model scores and planet candidate confirmation or rejection, suggesting that further evaluation of our high-scoring candidates is very likely to produce confirmed planets.}

We have provided the community with what we consider to be the most likely planetary candidates { from Kepler and TESS} to be confirmed.  Perhaps our most important contribution, however, is to demonstrate that the combination of instrument-specific datasets for training can yield an instrument-agnostic model, capable of generalization both inside and beyond the given instruments.  { On the cusp of the launch of Roman, we suggest that our simple model can be implemented to effectively prioritize exoplanet follow-up for confirmation.}


\section{Acknowledgements}
The authors thank N. Schanche for the helpful discussions and comments that have helped to improve this manuscript. {The authors also thank the anonymous referee for a review that allowed for substantial improvement of the paper.}  The material is based upon work supported by NASA under award number 80GSFC21M0002. SL is grateful to Jay Friedlander from NASA-GSFC for helping with the schematic in Fig 1.

\clearpage
\bibliography{sample631}{}
\bibliographystyle{aasjournal}

\end{document}